\documentclass[journal,comsoc,twoside]{IEEEtran}
\IEEEoverridecommandlockouts

\usepackage[T1]{fontenc}

\providecommand{\pgfsyspdfmark}[3]{}
\usepackage[utf8]{inputenc}
\usepackage{layouts}
\usepackage[normalem]{ulem}
\usepackage{xspace}
\usepackage{cite}
\usepackage{multirow}
\usepackage{setspace}
\usepackage{tikz}
\usepackage{xcolor}
\usepackage{makecell}
\usepackage[hyphens]{url}
\usepackage{graphicx}
\usepackage[binary-units=true,mode=text]{siunitx}
\usepackage{wasysym}
\usepackage{fancyhdr}

\makeatletter
\let\iint=\@undefined
\let\iiint=\@undefined
\let\iiiint=\@undefined
\makeatother

\usepackage[inline]{enumitem}
\setlist[itemize]{label=\textbullet,topsep=0pt,leftmargin=1em}
\setlist[enumerate]{label=(\alph*)}

\usepackage{listings}
\newcommand{\name}{CoinPrune\xspace}

\newcommand{\ie}{i.e.,\xspace}
\newcommand{\eg}{e.g.,\xspace}
\newcommand{\cf}{cf.\xspace}
\newcommand{\etal}{et al.\xspace}

\newcommand{\code}[1]{\texttt{\small#1}\normalsize}

\newcommand{\tabgod}{\CIRCLE}
\newcommand{\tabmed}{\LEFTcircle}
\newcommand{\tabbad}{\Circle}

\DeclareSIUnit\usd{USD}
\DeclareSIUnit\bitcoin{BTC}
\DeclareSIPrefix\billion{billion\ }{9}
\DeclareSIUnit{\million}{\text{m}}
\DeclareSIUnit{\mil}{\text{m}}

\usepackage[hidelinks]{hyperref}
\newcommand\copyrighttext{%
    \scriptsize%
    This is the author’s version of an article that has been published in IEEE Transactions on Network and Service Management (Volume: 18, Issue: 3, Sept. 2021).\\
    Changes were made to this version by the publisher prior to publication.
    The final version of record is available at \href{http://doi.org/10.1109/TNSM.2021.3073270}{http://doi.org/10.1109/TNSM.2021.3073270}.\\
    \textcopyright 2021 IEEE.
    Personal use is permitted. For any other purposes, permission must be obtained from the IEEE by emailing pubs-permissions@ieee.org.
}%
\IEEEoverridecommandlockouts

\fancypagestyle{IEEEtitlepagestyle}{%
    \fancyhf{}%
    \fancyfoot[c]{\copyrighttext{}}
}
\fancypagestyle{otherpages}{%
    \fancyhf{}%
    \fancyhead[RE,LO]{{\footnotesize MATZUTT \textit{et al.}: COINPRUNE: SHRINKING BITCOIN'S BLOCKCHAIN RETROSPECTIVELY}}
    \fancyhead[RO,LE]{{\footnotesize\thepage}}
    \fancyfoot[c]{\copyrighttext{}}
}
\newcommand{\btcstatAsOf}{Sep~29, 2020\xspace}

\newcommand{\btcstatChainsize}{\SI[round-mode=places,round-precision=0,per-mode=symbol]{280.6075886506587}{\gibi\byte}\xspace}
\newcommand{\btcstatGrowthrate}{\SI[round-mode=places,round-precision=0,per-mode=symbol]{142.7949284159848}{\mebi\byte\per day}\xspace}
\newcommand{\btcstatEvalSynctimeMaxVanilla}{\SI[round-mode=places,round-precision=0,per-mode=symbol]{6.745272333333333}{\hour}\xspace}
\newcommand{\btcstatEvalSynctimeMaxVanillaAbstract}{\SI[round-mode=places,round-precision=0,per-mode=symbol]{6.745272333333333}{\hour}\xspace}
\newcommand{\btcstatEvalSynctimeMaxCompaction}{\SI[round-mode=places,round-precision=0,per-mode=symbol]{51.475303333333336}{\minute}\xspace}
\newcommand{\btcstatEvalSynctimeMaxCompactionAbstract}{\SI[round-mode=places,round-precision=0,per-mode=symbol]{51.475303333333336}{\minute}\xspace}

\newcommand{\btcstatEvalTrafficRecvMaxVanillaAbstract}{\SI[round-mode=places,round-precision=0,per-mode=symbol]{270.66266648732125}{\gibi\byte}\xspace}

\newcommand{\btcstatEvalTrafficRecvMaxCompactionAbstract}{\SI[round-mode=places,round-precision=0,per-mode=symbol]{5.507908079773188}{\gibi\byte}\xspace}
\newcommand{\btcstatEvalTrafficBothMaxVanilla}{\SI[round-mode=places,round-precision=2,per-mode=symbol]{270.7015581486747}{\gibi\byte}\xspace}

\newcommand{\btcstatEvalTrafficBothMaxCompaction}{\SI[round-mode=places,round-precision=2,per-mode=symbol]{5.510472533851862}{\gibi\byte}\xspace}
\newcommand{\btcstatEvalTrafficMeanReduction}{\SI[round-mode=places,round-precision=2,per-mode=symbol]{92.98796482830397}{\percent}\xspace}
\newcommand{\btcstatEvalTrafficMeanReductionHighlevel}{\SI[round-mode=places,round-precision=0,per-mode=symbol]{92.98796482830397}{\percent}\xspace}
\newcommand{\btcstatEvalSynctimeMeanReduction}{\SI[round-mode=places,round-precision=2,per-mode=symbol]{77.03162841601736}{\percent}\xspace}
\newcommand{\btcstatEvalSynctimeMeanReductionHighlevel}{\SI[round-mode=places,round-precision=0,per-mode=symbol]{77.03162841601736}{\percent}\xspace}

\newcommand{\btcstatEvalStorageMaxVanillaHighlevel}{\SI[round-mode=places,round-precision=0,per-mode=symbol]{312.2071644347161}{\gibi\byte}\xspace}

\newcommand{\btcstatEvalStorageMaxCompactionHighlevel}{\SI[round-mode=places,round-precision=0,per-mode=symbol]{9.678908397443593}{\gibi\byte}\xspace}
\newcommand{\btcstatEvalStorageMaxSave}{\SI[round-mode=places,round-precision=2,per-mode=symbol]{302.5282560372725}{\gibi\byte}\xspace}

\newcommand{\btcstatEvalStorageMeanReduction}{\SI[round-mode=places,round-precision=2,per-mode=symbol]{86.9762398594862}{\percent}\xspace}
\newcommand{\btcstatEvalStorageMeanReductionHighlevel}{\SI[round-mode=places,round-precision=0,per-mode=symbol]{86.9762398594862}{\percent}\xspace}
\newcommand{\btcstatEvalStorageHeaderchainSize}{\SI[round-mode=places,round-precision=2,per-mode=symbol]{76.2942476272583}{\mebi\byte}\xspace}
\newcommand{\btcstatEvalStorageHeaderchainPerBlock}{\SI[round-mode=places,round-precision=2,per-mode=symbol]{133.33386166666668}{\byte}\xspace}

\newcommand{\btcstatEvalStorageMeanChaintailSizeLater}{\SI[round-mode=places,round-precision=2,per-mode=symbol]{1.0941895573799099}{\gibi\byte}\xspace}

\newcommand{\btcstatEvalStorageAntiContentsOverhead}{\SI[round-mode=places,round-precision=2,per-mode=symbol]{18.722803051644732}{\percent}\xspace}
\newcommand{\btcstatEvalStorageAntiContentsSizeVanilla}{\SI[round-mode=places,round-precision=2,per-mode=symbol]{3.898943672887981}{\gibi\byte}\xspace}
\newcommand{\btcstatEvalStorageAntiContentsSizeObfuscated}{\SI[round-mode=places,round-precision=2,per-mode=symbol]{4.628935217857361}{\gibi\byte}\xspace}

\newcommand{\btcstatEvalStorageAntiContentsFractionObfuscated}{\SI[round-mode=places,round-precision=1,per-mode=symbol]{99.24972551633849}{\percent}\xspace}

\newcommand{\btcstatEvalStorageOpreturnSize}{\SI[round-mode=places,round-precision=2,per-mode=symbol]{9.19904546905309}{\gibi\byte}\xspace}
\newcommand{\btcstatEvalStorageOpreturnAvgSize}{\SI[round-mode=places,round-precision=0,per-mode=symbol]{230.6987888390905}{\byte}\xspace}
\newcommand{\btcstatEvalStorageOpreturnNumber}{\SI[round-mode=places,round-precision=1,per-mode=symbol]{45.201755999999996}{\mil}\xspace}

\begin{document}

\bstctlcite{IEEEexample:BSTcontrol}

\title{\name: Shrinking Bitcoin's Blockchain Retrospectively}

\author{%
Roman Matzutt,~Benedikt Kalde,~Jan Pennekamp,~Arthur Drichel,~Martin Henze,~and Klaus Wehrle
\thanks{Roman Matzutt, Benedikt Kalde, Jan Pennekamp, and Klaus Wehrle are with the Chair of Communication and Distributed Systems, RWTH Aachen University, Aachen, Germany (e-mail: \{lastname\}@comsys.rwth-aachen.de).}%
\thanks{Arthur Drichel is with the IT Security Research Group, RWTH Aachen University, Aachen, Germany (e-mail: drichel@itsec.rwth-aachen.de).}%
\thanks{Martin Henze is with the Cyber Analysis \& Defense Department, Fraunhofer FKIE, Wachtberg, Germany (e-mail: martin.henze@fkie.fraunhofer.de).}}

\markboth{Matzutt \MakeLowercase{\textit{et al.}}: \name: Shrinking Bitcoin's Blockchain Retrospectively}%
{Matzutt \MakeLowercase{\textit{et al.}}: \name: Shrinking Bitcoin's Blockchain Retrospectively}

\maketitle

\begin{abstract}

Popular cryptocurrencies continue to face serious scalability issues due to their ever-growing blockchains.
Thus, modern blockchain designs began to prune old blocks and rely on recent snapshots for their bootstrapping processes instead.
Unfortunately, established systems are often considered incapable of adopting these improvements.
In this work, we present \emph{\name}, our block-pruning scheme with full Bitcoin compatibility, to revise this popular belief.
\name bootstraps joining nodes via snapshots that are periodically created from Bitcoin's set of unspent transaction outputs (UTXO set).
Our scheme establishes trust in these snapshots by relying on \name-supporting miners to \emph{mutually reaffirm} a snapshot's correctness on the blockchain.
This way, snapshots remain trustworthy even if adversaries attempt to tamper with them.
Our scheme maintains its retrospective deployability by relying on positive feedback only, \ie blocks containing invalid reaffirmations are not rejected, but invalid reaffirmations are outpaced by the benign ones created by an honest majority among \name-supporting miners.
Already today, \name reduces the storage requirements for Bitcoin nodes by two orders of magnitude, as joining nodes need to fetch and process only \btcstatEvalTrafficRecvMaxCompactionAbstract instead of \btcstatEvalTrafficRecvMaxVanillaAbstract of data in our evaluation, reducing the synchronization time of powerful devices from currently \btcstatEvalSynctimeMaxVanillaAbstract to \btcstatEvalSynctimeMaxCompactionAbstract, with even larger potential drops for less powerful devices.
\name is further aware of higher-level application data, \ie it conserves otherwise pruned application data and allows nodes to obfuscate objectionable and potentially illegal blockchain content from their UTXO set and the snapshots they distribute.

\end{abstract}

\vspace{-0.3em}

\vspace{-0.6em}

\section{Introduction}%
\label{sec:introduction}

\IEEEPARstart{K}{ey} to the success of cryptocurrencies such as Bitcoin~\cite{2008_nakamoto_bitcoin} is the blockchain, an immutable append-only ledger of financial transactions.
Using the blockchain, all nodes can independently verify the transaction history, allowing mutually distrusting peers to establish consensus about the correctness of those transactions.
This consensus also enables, for instance, audit systems~\cite{proof_of_existence,namecoin,2017_henze_dcam}, transparency overlays~\cite{2016_chase_transparency_overlays,2017_nikitin_chainiac}, bootstrapping anonymity services~\cite{2020_matzutt_anonboot}, and smart contracts~\cite{2016_wood_ethereum}.

However, besides all benefits blockchain systems face serious scalability challenges, \eg limited transaction throughput, high payment verification delays~\cite{2016_croman_blockchain_scalability}, and, most importantly, ever-growing blockchain sizes:
Bitcoin's blockchain exhibits a size of \btcstatChainsize with a recent average growth rate of \btcstatGrowthrate as of \btcstatAsOf~\cite{2011_blockchain_com_charts}.
Furthermore, the unintended utilization of Bitcoin's blockchain as a general-purpose data storage~\cite{2016_matzutt_poster,2018_matzutt_contents,2017_sward_contents} has put a permanent burden onto the system and its users, as 
\begin{enumerate*}
    \item such misuse typically bloats the set of unspent transaction outputs (UTXO set) with entries that are never spendable and
    \item objectionable content can irrevocably be engraved into the blockchain and is subsequently distributed to all nodes~\cite{2018_matzutt_thwarting}.
\end{enumerate*}
Large blockchain sizes and the presence of objectionable blockchain content cause individual nodes to \emph{prune} older blockchain data~\cite{2015_bitcoin_v0_11_pruning}, \ie older payment flows that have been superseded by newer ones, or \emph{locally erase} UTXOs that hold unwanted content~\cite{2019_florian_erase} at the cost of becoming dependent on other nodes for transaction validation.

While such decisions to prune or erase local blockchain data are rational from the users' perspective, they harm the overall network health:
To root trust in Bitcoin's current state, new nodes need to obtain and revalidate \emph{all} blockchain data, including all now-obsolete data, from independent sources.
Addressing this inherent conflict of interests, alternative designs~\cite{2014_bruce_mini_blockchain,2016_poelstra_mimblewimble,2016_chepurnoy_rollerchain,2019_schoenfeld_pascal} proposed \emph{snapshot-based synchronization}, where new nodes do not verify all blockchain data but instead rely on a recent snapshot of the blockchain's state.
While these efforts solve scalability challenges for new blockchain systems, they do not readily carry over to existing and well-established blockchain systems, such as Bitcoin, where significant changes have to be adopted by a majority of nodes~\cite{2017_morgan_scaling_debate}.
Consequently, there is a need to extend \emph{existing} blockchain systems with pruning capabilities in a secure and trustworthy manner without the need for significant changes of the underlying protocol.

\textbf{Our Contributions.}
We propose \emph{\name\footnote{Research prototype available at \url{https://github.com/COMSYS/coinprune}\\Snapshots available at \url{https://coinprune.comsys.rwth-aachen.de}}}, a block-pruning scheme that is fully compatible with Bitcoin and can be adopted immediately by any subset of nodes without changing Bitcoin's consensus rules.
At its core, \name provides a \emph{trustworthy} and fully distributed snapshot-based bootstrapping process based on Bitcoin's UTXO set, which is significantly smaller than Bitcoin's full blockchain.
Consequently, \name drastically unburdens the whole Bitcoin network as joining nodes require only the small snapshots for bootstrapping, and therefore established nodes can prune obsolete data \emph{without} impeding the overall network health.
In summary, we make the following contributions in this paper:

\begin{enumerate}[left=0pt,labelsep=3pt,label=\arabic*)]

\item We \emph{comprehensively survey} existing approaches to decrease the storage requirements and synchronization times of blockchain systems, concluding that they either are inefficient, insecure, or not deployable to existing systems.

\item We present \name, a snapshot-based pruning scheme that is fully compatible with Bitcoin.
\name establishes trust because \name-supporting miners \emph{periodically and independently reaffirm} the snapshots' correctness by cryptographically tying the current snapshot to the blockchain.
Thus, assuming a sufficient level of \name support, joining nodes may trust the honest network majority to verify snapshots before relying on them.
As legacy nodes ignore rather than reject reaffirmations, \name can be deployed mid-operation via a velvet fork~\cite{2017_kiayias_velvet_forks,2018_zamyatin_velvet_forks}.

\item \name further renders objectionable content in the UTXO set (and thus its snapshots) harmless by \emph{obfuscating} values to make the content inaccessible without losing the corresponding transactions' verifiability.
    Simultaneously, \name preserves short chunks of non-financial data from \code{OP\_RETURN} transaction outputs, which are essential for higher-level applications such as audit systems~\cite{proof_of_existence,namecoin,2017_henze_dcam}, in an additional \emph{application data storage}.

\item Our evaluation shows that \name reduces disk space utilization for full nodes by \btcstatEvalStorageMeanReductionHighlevel while still enabling new nodes to synchronize.
    Pruning improves the synchronization performance drastically:
Network traffic is reduced by \btcstatEvalTrafficMeanReductionHighlevel and synchronization times drop from \btcstatEvalSynctimeMaxVanilla to \btcstatEvalSynctimeMaxCompaction on powerful devices with even larger drops on less powerful devices.
Further, obfuscating snapshots keeps current snapshot sizes at \btcstatEvalStorageAntiContentsSizeObfuscated, and retaining all \code{OP\_RETURN} data together with relevant metadata requires \btcstatEvalStorageOpreturnSize.
\end{enumerate}

A preliminary version of this paper appears in the proceedings of IFIP Networking 2020~\cite{2020_matzutt_coinprune}.
We extend and improve upon our previous work in the following ways:
First, we extend \name to \emph{obfuscate almost all objectionable content}~\cite{2018_matzutt_contents} to make such content inaccessible without hurting transaction verifiability.
Second, we \emph{extend our security discussion} by simulatively assessing an adversary's success chance to reaffirm an invalid snapshot.
Third, we introduce an application data storage for the \emph{preservation of application-level data}.
Fourth, we \emph{extend our discussion of related work} to also cover new approaches and work that does not primarily target synchronization scalability.
Finally, we \emph{updated our performance evaluation} with newer data from Bitcoin's blockchain.

\textbf{Paper Structure.}
Section~\ref{sec:background} provides the necessary background on Bitcoin.
In Section~\ref{sec:motivation}, we discuss the negative impacts of growing blockchains.
We then extensively survey related work in Section~\ref{sec:sota} and identify requirements for secure block-pruning schemes based on this survey in Section~\ref{sec:requirements}.
Section~\ref{sec:overview} provides an overview of \name's design.
We then elaborate on its Bitcoin-retrofittable block-pruning scheme in Section~\ref{sec:pruning}, its handling of application-level data in Section~\ref{sec:data-semantics}, and its integration into Bitcoin in Section~\ref{sec:integration}.
We discuss the security of \name in Section~\ref{sec:security} and evaluate its performance in Section~\ref{sec:eval}.
Section~\ref{sec:conclusion} concludes this paper.

\section{Bitcoin Overview}%
\label{sec:background}

We begin with a primer on Bitcoin before describing its transaction management, its cache of relevant transaction outputs, implications of non-financial transaction data, the deployment of consensus updates, and the bootstrapping process.

\textbf{Bitcoin Primer.}
Bitcoin's~\cite{2008_nakamoto_bitcoin} blockchain provides a public and immutable append-only ledger of financial transactions to prevent the double-spending of coins within an untrusted peer-to-peer (P2P) network.
Bitcoin establishes this ledger by bundling pending transactions in cryptographically interlinked, hard-to-create blocks.
At its core, the P2P network consists of two special types of nodes.
First, \emph{full nodes} maintain the blockchain by locally validating all pending transactions and proposed new blocks and by discarding any incorrect data.
Second, \emph{miners} invest their hardware resources to find new blocks while solving a proof-of-work (PoW) puzzle in exchange for freshly minted bitcoins as a reward.
Modifying blocks at a later point becomes increasingly hard as it requires recomputing all subsequent blocks to keep their chaining intact.
On a technical level, Bitcoin blocks consist of a \emph{header} and a set of transactions.
The \SI{80}{\byte}-long header contains the block's version, the hash value of the block's predecessor for chaining blocks together, the root of a Merkle tree over the block's transactions to cryptographically tie them to the block header, the time the block was mined, and the miner's PoW.

\textbf{Transaction Management.}
Each transaction transfers previously received or newly minted bitcoins to one or more receivers via individual \emph{transaction outputs}.
To prevent double-spending, full nodes have to verify any claimed coin ownership for all pending transactions using a dedicated scripting language~\cite{2010_bitcoin_script}.
Each transaction output defines a spending condition using this script language, which typically sends coins to a specific \emph{Bitcoin address} that corresponds to a cryptographic key pair.
A user can prove that she owns the coins associated with a transaction output by providing the data required to satisfy the spending condition in a \emph{transaction input} of a subsequent transaction.
For efficiency reasons, all nodes keep track of the current \emph{set of unspent transaction outputs (UTXO set)}.
Thereby, nodes can discard pending transactions that attempt to spend non-existing or already spent bitcoins.
Notably, by default, full nodes prune all spent transactions, \ie transactions that do not contribute to the UTXO set anymore, from their transaction index.
Nevertheless, these nodes retain a full copy of all blocks for bootstrapping new nodes.

\textbf{UTXO Set Layout.}
When processing incoming transactions, each node adds all those transaction outputs to the UTXO set that are not \emph{provably} unspendable.
For instance, the standard means of transferring funds in Bitcoin is via pay-to-public-key-hash (P2PKH) transaction outputs, which implement the aforementioned sending of funds to a specific Bitcoin address.
The spending condition associated with a P2PKH transaction output requires the designated spender to present the public key matching a cryptographic hash value given in the P2PKH script and a digital signature created using the corresponding private key.
However, nodes receiving a transaction have no way to verify that this private key is indeed known to any user~\cite{2018_matzutt_thwarting}.
Therefore, nodes are forced to include potentially unspendable transaction outputs in their UTXO set, which would then reside there forever.
For a more succinct representation of the UTXO set, nodes \emph{compress} simple and recurring scripts within the UTXOs, \eg P2PKH scripts that always have the same \SI{25}{\byte}-long pattern, with only the \SI{20}{\byte}-long hash value changing.
Consequently, nodes only store the hash value in their UTXO set and decompress the original transaction output on the fly when validating a transaction spending this specific UTXO.
In total, Bitcoin Core distinguishes six compressible script types (P2PKH, P2SH, and P2PK with four different types of public keys) and stores any other potentially spendable UTXO in uncompressed form.

\textbf{Non-Financial Data.}
Besides the aforementioned financial transactions, users can also permanently store non-financial data on Bitcoin's blockchain.
Even though the technical methods for inserting such data vary~\cite{2018_matzutt_contents}, content is typically inserted in one of the following two ways.
First, using a special script starting with the \code{OP\_RETURN} operation in one output allows the user to augment her transaction with up to \SI{80}{\byte} of arbitrary data.
As such transaction outputs are provably unspendable, nodes do not add them to their UTXO set.
Second, other users manipulate the mutable values of standard financial transactions, \eg hash values within P2PKH transaction outputs, as an unintended means to insert arbitrary data at a higher capacity~\cite{2018_matzutt_contents}.
As this manipulation cannot be detected reliably and the affected transaction outputs cannot be determined to be unspendable~\cite{2018_matzutt_thwarting}, such transaction outputs are added to the UTXO set and are likely to remain there forever.
Consequently, any objectionable content may not only enter the immutable blockchain but also reside in the UTXO set indefinitely.
We hence refer to this phenomenon as \emph{pollution} of the UTXO set for the remainder of this paper.

\textbf{Updates of Consensus Rules.}
Bitcoin must tolerate and resolve blockchain \emph{forks}, \ie situations in which the blockchain diverges into more than one potential path forward due to its distributed nature.
Forks can occur either accidentally, due to the concurrent mining process, or with intent.
Accidental forks dissolve naturally since Bitcoin nodes only consider the fastest-growing branch to be valid, and one branch is highly likely to grow faster than other competing branches.
Intentional forks are used to \emph{update} existing consensus rules, and they are traditionally categorized as either \emph{hard forks} or \emph{soft forks}~\cite{2018_zamyatin_velvet_forks}.
While hard forks introduce protocol-breaking changes, \eg altered block structures, soft forks aim to remain backward-compatible with clients following older consensus rules~\cite{2018_zamyatin_velvet_forks}.
Both paradigms can incur \emph{permanent} blockchain forks depending on whether the majority of nodes accept or reject the proposed changes~\cite{2018_zamyatin_velvet_forks}.
Contrarily, multiple works recently investigated \emph{velvet forks}~\cite{2017_kiayias_velvet_forks,2018_zamyatin_velvet_forks}, which aim for the \emph{gradual introduction} of new features without creating permanent forks.
This type of fork augments upgraded blocks such that they remain valid to legacy nodes, but updated nodes process them according to the changed protocol.

\textbf{Initial Synchronization.}
When a node first joins the Bitcoin network, it needs to obtain its individual view on Bitcoin's current state of consensus, \ie the UTXO set that results from the blockchain path containing the most PoW.
To keep this process fully decentralized and independent from trusted nodes, each node initially establishes eight outgoing connections to random other nodes, called \emph{neighbors}, and downloads the complete blockchain from them.
Due to the separation of headers and transactions, nodes first fetch the \emph{headerchain}, \ie the chain of block headers, and simultaneously request full blocks, \ie the corresponding transactions.
While receiving the data, the joining node verifies its correctness by
\begin{enumerate*}
    \item{verifying the blockchain's cryptographic links back to the hard-coded genesis block,}
    \item{keeping track of the amount of performed PoW to remain on the currently valid blockchain,}
    \item{validating the transaction sets tied to each block, and}
    \item{verifying and replaying all transactions to obtain an up-to-date UTXO set.}
\end{enumerate*}
Even though the headerchain and the UTXO set are sufficient to process new blocks, the nodes keep a full blockchain copy by default to be able to bootstrap other nodes.

\section{Impact of Long-Term Blockchain Utilization}%
\label{sec:motivation}

Blockchains continuously grow in size by design and thus eventually reach prohibitive sizes.
For instance, Bitcoin's blockchain currently has a size of \btcstatChainsize with an average growth rate of \btcstatGrowthrate as of \btcstatAsOf~\cite{2011_blockchain_com_charts}.
This worrying trend severely impacts the scalability of the overall system.
In addition to increasing \emph{storage requirements}, which already exclude whole device classes such as smartphones from operating a full node, there are additional influences on the \emph{required bandwidth}, \emph{processing costs}, and \emph{synchronization times} of joining nodes.
Finally, the immutability property may cause pollution with unwanted data over time.

\textbf{Storage Requirements.}
To retain a decentralized consensus within its network, Bitcoin requires that enough independent nodes maintain a full blockchain copy to be able to bootstrap joining nodes (\cf~Section~\ref{sec:background}).
However, storing hundreds of Gigabytes of past blockchain data is irrational for individual node operators and prohibitive for storage-constrained devices.
Consequently, such devices cannot act as full nodes, and users have to accept weakened security guarantees.

\textbf{Bandwidth Requirements.}
During the initial synchronization, each joining node must obtain a full blockchain copy.
Current blockchain sizes already require good Internet connectivity for both the joining node and its neighbors, potentially causing increased initial synchronization times for joining nodes.
Furthermore, such requirements also put an additional burden onto existing nodes as serving new nodes consumes resources that could otherwise be used for other tasks, \eg gossiping pending transactions or newly mined blocks.

\textbf{Processing Costs.}
In addition to downloading the blockchain, joining nodes also need to verify the blockchain's integrity and locally replay every single transaction to build the UTXO set.
This process consumes excessive amounts of computation power for joining nodes~\cite{2015_bitcoin_fullnode,2019_geniar_bitcoin_performance,2020_lopp_bitcoin_performance}.
In particular, large numbers of obsolete transactions that do not contribute to the UTXO set anymore waste valuable resources.

\textbf{Synchronization Time.}
High bandwidth requirements and high processing costs combined cause prolonged synchronization times.
While benchmarks report synchronization times of about \num{5}--\SI{8}{\hour} between 2018 and 2020~\cite{2019_geniar_bitcoin_performance,2020_lopp_bitcoin_performance}, literature already highlighted this issue in 2016 when four days were required to synchronize Amazon EC2 nodes~\cite{2016_croman_blockchain_scalability}.
This problem aggravates over time as new blocks are added continuously.

\textbf{Blockchain Pollution.}
Blockchain immutability ensures that agreed-upon events or transactions cannot be altered at a later point.
However, undesired data, such as manipulated transactions holding illicit content, cannot be removed retrospectively either, and such data also may persistently bloat the UTXO set, leading to UTXO set pollution (\cf~Section~\ref{sec:background}).
On the one hand, a bloated UTXO set implies further scalability issues during the verification of transactions~\cite{2019_bartoletti_metadata}.
On the other hand, illicit blockchain content is known to imply legal risks for blockchain users~\cite{2018_matzutt_contents}.

In summary, the ever-growing blockchain implies strong scalability issues for both joining and established nodes, and the presence of illicit blockchain content presents blockchain nodes with further security challenges.
Established nodes are even punished for altruistically serving the full blockchain to help joining nodes synchronize.
Especially systems such as Bitcoin, which experience a long-term utilization, suffer from these problems already today.
In the following, we thus survey to which extent (newly proposed) systems tackle these issues.

\begin{table*}[t]
    \centering
    \caption{Qualitative Comparison of Approaches Improving Storage Requirements and Initial Synchronization}
    \vspace{-0.7em}
    \setlength{\tabcolsep}{5pt}
    \scriptsize
    \begin{tabular}{p{0.0cm}l|l||c|c|c|c|c|c|c|c|c}
        &
        \textbf{Name} &
        \textbf{Approach} &
        \textbf{\makecell{Reduce \\ Processing}} &
        \textbf{\makecell{Reduce \\ Traffic}} &
        \textbf{\makecell{Reduce \\ Storage}} &
        \textbf{\makecell{Sync. \\ Time}} &
        \textbf{\makecell{Maintain \\ Security}} &
        \textbf{\makecell{Network \\ Health}} &
        \textbf{\makecell{Server \\ Burden}} &
        \textbf{\makecell{Comple- \\ teness}} &
        \textbf{\makecell{Compat- \\ ibility}} \\
        \hline
        \multirow{9}{*}{\hspace{-0.1cm}\rotatebox{90}{Deployed Solutions}} &
        Hot Wallets~\cite{2012_bitcoin_hot_wallet} &
        Trust Delegation &
        \tabbad\ / \tabgod &
        \tabbad\ / \tabgod &
        \tabbad\ / \tabgod &
        \tabbad\ / \tabgod &
        \tabbad &
        \tabbad &
        \tabgod &
        \tabbad &
        \tabgod \\
        &
        Light Nodes~\cite{2018_bitcoin_light_nodes} &
        Trust Delegation &
        \tabbad\ / \tabgod &
        \tabbad\ / \tabgod &
        \tabbad\ / \tabgod &
        \tabbad\ / \tabgod &
        \tabmed &
        \tabbad &
        \tabgod &
        \tabbad &
        \tabgod \\
        &
        ``Ultimate Compression''~\cite{2012_reiner_ultimate_compression} &
        Trust Delegation &
        \tabbad\ / \tabgod &
        \tabbad\ / \tabgod &
        \tabbad\ / \tabgod &
        \tabbad\ / \tabgod &
        \tabbad &
        \tabgod &
        \tabbad &
        \tabgod &
        \tabgod \\
        &
        DB Improvements~\cite{2013_bitcoin_v0_8_0_db_change,2017_bitcoin_v0_15_utxo_changed} &
        Data Management &
        \tabmed &
        \tabbad &
        \tabbad &
        \tabmed &
        \tabgod &
        \tabgod &
        \tabgod &
        \tabgod &
        \tabgod \\
        &
        Index Pruning~\cite{2013_bitcoin_v0_8_0_db_change} &
        Data Management &
        \tabmed &
        \tabbad &
        \tabmed &
        \tabmed &
        \tabgod &
        \tabgod &
        \tabgod &
        \tabmed &
        \tabgod \\
        &
        Headers-first Download~\cite{2015_bitcoin_v0_10_synchronization} &
        Data Management &
        \tabbad &
        \tabbad &
        \tabbad &
        \tabmed &
        \tabgod &
        \tabgod &
        \tabgod &
        \tabgod &
        \tabmed \\
        &
        Assume-valid Blocks~\cite{2010_nakamoto_v0_3_2_checkpoints,2017_bitcoin_v0_14_assumevalid} &
        Skip Verification &
        \tabgod &
        \tabbad &
        \tabbad &
        \tabmed &
        \tabgod &
        \tabgod &
        \tabgod &
        \tabgod &
        \tabgod \\
        &
        Block Pruning~\cite{2015_bitcoin_v0_11_pruning} &
        Simple Block Pruning &
        \tabbad &
        \tabbad &
        \tabgod &
        \tabbad &
        \tabgod &
        \tabbad &
        \tabgod &
        \tabmed &
        \tabgod \\
        &
        Ethereum Fast Sync~\cite{2015_szilagyi_ethereum_fast_sync} &
        State-based Sync. &
        \tabgod &
        \tabbad &
        \tabbad &
        \tabgod &
        \tabgod &
        \tabgod &
        \tabgod &
        \tabgod &
        \tabbad \\
        &
        AssumeUTXO~\cite{2019_obeirne_assumeutxo} &
        State-based Sync. &
        \tabbad &
        \tabbad &
        \tabbad &
        \tabgod &
        \tabmed &
        \tabgod &
        \tabgod &
        \tabgod &
        \tabgod \\
        \hline
        \multirow{10}{*}{\hspace{-0.1cm}\rotatebox{90}{Related Work}} &
        Selective Pruning~\cite{2018_palm_selective_pruning} &
        Simple Block pruning &
        \tabmed  &
        \tabmed  &
        \tabgod  &
        \tabmed  &
        \tabbad  &
        \tabgod  &
        \tabbad  &
        \tabgod  &
        \tabbad \\
        &
        Rollerchain~\cite{2016_chepurnoy_rollerchain} &
        State-based Sync. &
        \tabgod &
        \tabgod &
        \tabgod &
        \tabgod &
        \tabgod &
        \tabgod &
        \tabmed &
        \tabmed &
        \tabbad \\
        &
        Marsalek \etal~\cite{2019_marsalek_compression} &
        State-based Sync. &
        \tabbad\ / \tabgod &
        \tabbad\ / \tabgod &
        \tabbad\ / \tabgod &
        \tabgod &
        \tabgod &
        \tabmed &
        \tabmed &
        \tabmed &
        \tabbad \\
        &
        Mini Blockchain Scheme~\cite{2014_bruce_mini_blockchain} &
        Balance-based Sync. &
        \tabgod &
        \tabgod &
        \tabgod &
        \tabgod &
        \tabgod &
        \tabmed &
        \tabmed &
        \tabbad &
        \tabbad \\
        &
        Mimblewimble~\cite{2016_poelstra_mimblewimble} &
        Balance-based Sync. &
        \tabbad &
        \tabmed &
        \tabmed &
        \tabbad &
        \tabgod &
        \tabgod &
        \tabgod &
        \tabbad &
        \tabbad \\
        &
        Pascal~\cite{2019_schoenfeld_pascal} &
        Balance-based Sync. &
        \tabgod &
        \tabgod &
        \tabgod &
        \tabgod &
        \tabmed &
        \tabbad &
        \tabgod &
        \tabmed &
        \tabbad \\
        &
        Vault~\cite{2019_leung_vault} &
        Balance-based Sync. &
        \tabgod &
        \tabgod &
        \tabgod  &
        \tabgod  &
        \tabgod  &
        \tabgod  &
        \tabgod  &
        \tabmed  &
        \tabbad \\
        &
        FlyClient~\cite{2020_buenz_flyclient} &
        Commitment-based &
        \tabbad\ / \tabgod &
        \tabbad\ / \tabgod &
        \tabbad\ / \tabgod &
        \tabbad\ / \tabgod &
        \tabgod &
        \tabbad &
        \tabgod &
        \tabbad &
        \tabbad \\
        &
        TICK~\cite{2020_zhang_tick} &
        Commitment-based &
        \tabbad\ / \tabgod &
        \tabbad\ / \tabgod &
        \tabbad\ / \tabgod &
        \tabbad\ / \tabgod &
        \tabgod &
        \tabbad &
        \tabgod &
        \tabbad &
        \tabbad \\
        &
        MiniChain~\cite{2020_chen_minichain} &
        Commitment-based &
        \tabgod &
        \tabgod &
        \tabgod &
        \tabgod &
        \tabgod &
        \tabgod &
        \tabgod &
        \tabbad &
        \tabbad \\
		\hline
        &
        \name (our approach) &
        State-based Sync. &
        \tabgod &
        \tabgod &
        \tabgod &
        \tabgod &
        \hphantom{*}\tabgod* &
        \hphantom{*}\tabgod* &
        \tabmed &
        \tabmed &
        \tabgod \\
    \end{tabular}

    \vspace{0.3em}
    \hspace{3.4em}{%
    \itshape
    $\Box\, /\,\Box$: Distinction Full Nodes / Light Nodes
    \hfill
    $*$: Dependent on honest majority among adopters (\cf~Section~\ref{sec:security})
    }\hspace{2.4em}
    \label{tab:comparison}
    \vspace{-2.4em}
\end{table*}

\vspace{-0.2em}
\section{The Current State of Blockchain Pruning}%
\label{sec:sota}

Learning from the observed scalability issues, developers of new blockchain systems tackled these challenges from different perspectives.
In this section, we survey current state-of-the-art measures deployed to existing systems and alternative blockchain designs that focus on reducing storage requirements and improving the bootstrapping.
Finally, we provide an overview of further related work that does not primarily aim to improve the bootstrapping process.

\vspace{-0.5em}
\subsection{Survey Criteria and Methodology}%
\label{sec:sota:criteria}

We qualitatively assess the applicability and effectiveness of related approaches based on
\begin{enumerate*}
    \item{their scalability improvements,}
    \item{their capability to maintain sufficient security levels,}
    \item{their impact on the overall network,}
    \item{their potential impact on blockchain queryability, and}
    \item{their compatibility with already established public blockchains, such as Bitcoin.}
\end{enumerate*}

We assess \emph{scalability improvements} by considering processing, traffic, and storage improvements separately and, from this, we infer the impact on synchronization times for joining nodes.
We resort to a qualitative assessment of the presented approaches, as most works do not provide comparable performance benchmarks.
Furthermore, we discuss to which extent the improvements impact their base systems' \emph{security} guarantees.
The \emph{impact on the overall network} engulfs consequences for the \emph{network health}, \ie its dependency on especially altruistic nodes and potential additional overheads imposed on established nodes.
Then, we assess how the proposed schemes may impact the \emph{blockchain queryability}, \eg the capability of querying now-obsolete transactions or augmented transactions such as Bitcoin's \code{OP\_RETURN} transactions.
Finally, we survey their \emph{compatibility} with already deployed blockchain systems.
We summarize our results in Table~\ref{tab:comparison}.

\vspace{-0.5em}
\subsection{Measures Deployed in Existing Blockchain Systems}%
\label{sec:sota:updates}

The increased popularity of cryptocurrencies forced their developers to tackle rising scalability issues.
We now discuss measures taken either locally by users or network-wide by blockchain developers.
Our discussion is based on the reference implementations (Bitcoin Core and Ethereum's \code{geth}, respectively) where appropriate.
Overall, we identify approaches based on \emph{trust delegation}, \emph{skipping verification} steps, \emph{improving data management} with the special case of \emph{block pruning}, and \emph{state-based synchronization}.

\textbf{Trust Delegation.}
Users rely on third parties to root trust in the blockchain's correctness if they cannot operate a full node, \eg when using a constrained device for issuing transactions.
Using \emph{hot wallets}~\cite{2012_bitcoin_hot_wallet}, users essentially outsource all fund management to a trusted third party, enabling the service provider to issue transactions on their behalf.
Similarly, \emph{light nodes}~\cite{2018_bitcoin_light_nodes} outsource blockchain verification to other full nodes, but they manage their wallet locally using simplified payment verification (SPV)~\cite{2008_nakamoto_bitcoin}.
These approaches vastly improve the performance of clients, only put a negligible burden on the full nodes, and they are actively used.
However, clients do not contribute positively to the overall network using these approaches as they only seize other nodes' resources.
Contrarily, trust-delegating nodes heavily rely on a backbone network of full nodes for both trust and relevant information and prohibit local verifiability.
The never-deployed Ultimate Compression scheme~\cite{2012_reiner_ultimate_compression} aimed at bootstrapping light nodes with the current UTXO set but requires full nodes to store and transmit a searchable representation of the UTXO set in addition to its full blockchain copy, putting an extra burden on the full nodes.
Furthermore, this scheme requires an additional blockchain to establish trust in the transmitted UTXO set.

\textbf{Improving Data Management.}
Increasing blockchain sizes necessitate optimized data management, either for looking up relevant information or for bootstrapping new nodes efficiently.
To this end, Bitcoin Core has historically changed its \emph{underlying database} system~\cite{2013_bitcoin_v0_8_0_db_change} and the internal layout of its UTXO set~\cite{2017_bitcoin_v0_15_utxo_changed}.
Furthermore, full nodes \emph{locally prune} obsolete entries from their \emph{transaction index}~\cite{2013_bitcoin_v0_8_0_db_change}.
While the nodes still persist the full raw blockchain data, such obsolete information is not queryable anymore.
Network-related optimizations mainly engulf a revised \emph{header-first download} of the blockchain~\cite{2015_bitcoin_v0_10_synchronization}.
Verifying the headerchain is sufficient to ensure the blockchain's integrity.
Since transactions can be decoupled from their block's headers (\cf~Section~\ref{sec:background}), nodes can now download and verify full blocks in parallel with only minor and local upgrading incompatibilities~\cite{2015_bitcoin_v0_10_synchronization}.
They can further \emph{limit their block-serving bandwidth}~\cite{2016_bitcoin_v0_12_limit_upload} and relay \emph{compact representations} of newly mined blocks, which avoids broadcasting known-but-pending transactions redundantly~\cite{2016_bitcoin_v0_13_segwit}.
However, the header-first download still requires nodes to obtain and process all blockchain data during their initial synchronization, and only improves the distribution of that data.

\textbf{Skipping Verification.}
Early on, Bitcoin's reference implementation started to avoid revalidating transactions from very old blocks.
Using hard-coded \emph{checkpoint blocks} at first~\cite{2010_nakamoto_v0_3_2_checkpoints}, Bitcoin has recently shifted to use configurable \emph{assumed-valid blocks}~\cite{2017_bitcoin_v0_14_assumevalid}.
The reasoning here is that invalid transactions would have been rejected by the network earlier, and thus older transactions with many confirmations are believed to be correct.
By skipping assumed-valid blocks, joining nodes can avoid the costly signature verification of large portions of the blockchain at negligible security risks.
However, joining nodes still download the complete blockchain to replay all transactions to create an up-to-date UTXO set.

\textbf{Simple Block Pruning.}
To counter increasing storage requirements, Bitcoin users have the option to completely \emph{prune raw blockchain data}~\cite{2015_bitcoin_v0_11_pruning} after a \emph{full} initial synchronization.
This step allows nodes to forget all obsolete blockchain data at the cost of its queryability.
In contrast to local index pruning, block pruning is detrimental to the network health as block-pruning nodes are incapable of bootstrapping new nodes.

\textbf{State-based Synchronization.}
While Bitcoin focuses on financial transactions, other cryptocurrencies, such as Ethereum~\cite{2016_wood_ethereum}, can also execute smart contracts.
Naturally, those cryptocurrencies have more complex state layouts as the full nodes need to keep track of every smart contract's state.
Consequently, Ethereum uses \emph{Fast Sync}~\cite{2015_szilagyi_ethereum_fast_sync}, which enables joining nodes to download a recent state and avoid replaying all blockchain data.
However, Ethereum still values the queryability of older data.
Thus, joining nodes also download and persist all blocks but they do not have to process them during their initial synchronization.
In contrast to Bitcoin's proposal for Ultimate Compression, Fast Sync remains secure since Ethereum, by default, cryptographically ties its current state to each block~\cite{2016_wood_ethereum}.
Hence, nodes can verify the correctness of their obtained state directly via Ethereum's blockchain.
Since other cryptocurrencies lack these header fields, Fast Sync is not immediately portable.
AssumeUTXO~\cite{2019_obeirne_assumeutxo} aims for a similar extension for Bitcoin by extending upon its assume-valid blocks.
Joining nodes use state-based synchronization based on a recent UTXO set to be operable early on, but then transition to a full background synchronization to retain full queryability.
Furthermore, AssumeUTXO relies on hard-coded cryptographic state identifiers as well as third-party servers for the state distribution, which affects its security.

\textbf{Takeaway.}
Developers have tackled the scalability issues of blockchain systems from different perspectives.
However, all approaches either have limited efficiency or questionable security properties, they are detrimental to the network health, or they are not readily portable to already established systems.

\vspace{-1em}
\subsection{Proposed Block-Pruning Schemes}%
\label{sec:sota:proposals}

The lacking efficiency of post-deployment measures motivated various \emph{alternative blockchain designs} that promise improved scalability measures.
We identify alternative designs that refine mere \emph{block-pruning schemes}, designs proposing \emph{state-based} or \emph{balance-based} synchronization, and \emph{commitment-based} designs aiming to further reduce or even remove the need for locally maintaining the current state.

\textbf{Simple Block Pruning.}
Palm \etal~\cite{2018_palm_selective_pruning} present a distributed block-pruning scheme for established nodes in permissioned blockchains, \ie blockchains jointly maintained by a fixed set of mutually known parties.
A dedicated pruning initiator defines a pruning algorithm that all nodes must execute to identify and prune now-irrelevant transactions in a way that other nodes can still retrieve all relevant data.
However, this approach focuses on permissioned blockchains and requires a dedicated initiator.
Hence, the approach is inapplicable to public settings, which are open to unknown or unauthenticated parties, both for security and compatibility reasons.

\textbf{State-based Synchronization.}
Similar to Ethereum's Fast Sync, and inheriting its advantages and disadvantages, Rollerchain~\cite{2016_chepurnoy_rollerchain} proposes a state-based bootstrapping process.
However, Rollerchain values performance over complete queryability, thereby significantly decreasing synchronization overheads as old information can be fully pruned.
Similarly, Marsalek \etal~\cite{2019_marsalek_compression} propose a state-based bootstrapping process based on Bitcoin, but the approach forfeits compatibility with Bitcoin by rejecting blocks with invalid states attached.

\textbf{Balance-based Synchronization.}
A special class of state-based block-pruning schemes simplifies the structure of what constitutes the state to allow for more efficient representations and updates~\cite{2014_bruce_mini_blockchain,2016_poelstra_mimblewimble,2019_schoenfeld_pascal,2019_leung_vault}.
Typically, these schemes only keep track of existing accounts and their balances.
The Mini-Blockchain scheme~\cite{2014_bruce_mini_blockchain} replaces Bitcoin's UTXO set with an account tree that is cryptographically tied to each mined block.
Joining nodes obtain the headerchain and a recent account tree to synchronize before fully processing a tail of full blocks to preserve PoW-based security.
However, the scheme expects established nodes to compute slices of the recent account tree on demand without elaborating how it ensures the availability of all required data to rewind the account tree accordingly.
Mimblewimble~\cite{2016_poelstra_mimblewimble} follows a similar approach but emphasizes confidential transactions at the cost of synchronization performance as joining nodes have to obtain and verify rangeproofs for unspent funds~\cite{2016_poelstra_mimblewimble}.
Through their balance-based approach, both schemes limit the expressiveness of transactions.
To overcome this limitation, Pascal~\cite{2019_schoenfeld_pascal} defines SafeBoxes as a replacement for mere account trees.
SafeBoxes permit the generation of a limited number of accounts per block and are designed to enable higher-layer applications.
However, the limited availability of account spots is conceptually detrimental to network health.
Finally, Vault~\cite{2019_leung_vault} builds upon Algorand~\cite{2017_gilad_algorand} to enable the distribution of fragments of recent states across the network to reduce the per-node storage requirements.
Therefore, Vault is inapplicable as an aid for existing, simpler cryptocurrencies.

\textbf{Commitment-based Improvements.}
Recently, multiple approaches proposed to extend the commitment to the current state within block headers and make state data queryable this way~\cite{2020_buenz_flyclient,2020_zhang_tick,2020_chen_minichain}.
FlyClient~\cite{2020_buenz_flyclient} reduces a client's overhead when relying on SPV by avoiding the need to obtain the full headerchain from other nodes to convince the client that its contact node relies on a valid blockchain.
FlyClient achieves this goal by maintaining a Merkle mountain range~\cite{2012_todd_mmr} over the headerchain.
A Merkle mountain range is an append-only variant of a Merkle tree that allows for efficient membership tests~\cite{2020_buenz_flyclient}.
In FlyClient, a light node uses this commitment in each block to probabilistically challenge full nodes about the correctness of their headerchain before accepting their blocks and engage in SPV~\cite{2020_buenz_flyclient}.
TICK~\cite{2020_zhang_tick} instead encodes the UTXO set as an AVL hash tree that commits to the UTXO set in a way that
\begin{enumerate*}
    \item changes to the UTXO set, \ie insertion and deletion of UTXOs when processing a new block, can efficiently be reflected based on an existing AVL hash tree and
    \item light nodes can efficiently verify the presence or absence of a specific UTXO based on an up-to-date AVL hash tree~\cite{2020_zhang_tick}.
\end{enumerate*}
Both approaches assume the presence of full nodes that store the full UTXO set but that can provide light nodes with a succinct proof to convince them of their honesty.
Contrarily, MiniChain~\cite{2020_chen_minichain} aims for a practical \emph{stateless blockchain} that overcomes the need for storing the full UTXO set for verification purposes. 
Based on prior work by Boneh \etal~\cite{2019_boneh_accumulators}, MiniChain uses RSA-based accumulators suited for the distributed setting to commit to the UTXO set in a similar way to TICK.
However, MiniChain requires fund spenders to provide an additional proof to convince validating nodes that the transaction only spends existing coins that have not previously been spent~\cite{2020_chen_minichain}.
Similarly to balanced-based pruning schemes, these commitments are, however, incompatible with already deployed cryptocurrencies.

\textbf{Takeaway.}
Alternative blockchain designs have shown that incorporating cryptographic ties to recent state objects are a promising means to establish trust in state-based blockchain synchronization processes.
However, extending existing systems with such capabilities immediately results in hard forks, which are difficult to deploy and thus highly debated~\cite{2017_morgan_scaling_debate}.

\subsection{Further Related Work}%
\label{sec:sota:related-work}

Other works that consider blockchain data management include analyses of blockchain data~\cite{2013_ron_transaction_analysis,2013_meiklejohn_anonymity_analysis,2017_bartoletti_opreturn,2016_matzutt_poster,2018_matzutt_contents,2017_sward_contents,2019_bartoletti_metadata} and the UTXO set~\cite{2018_delgado_segura_utxo}, lightweight payment schemes~\cite{2016_poon_lightning,2017_green_bolt}, approaches to prevent illicit content from being engraved into the blockchain~\cite{2017_ateniese_redactable_blockchain,2017_puddu_mu_chain,2018_matzutt_thwarting,2019_deuber_redactable_blockchain,2019_florian_erase,2019_dorri_mofbc}, and sharding approaches~\cite{2016_luu_elastico,2018_zamani_rapidchain,2018_kokoris_omniledger}.
In the following, we provide pointers to cover the research perspectives for this further related work.

\textbf{Previous Analyses.}
Bitcoin's blockchain and scalability properties have previously been analyzed from different perspectives.
Initial work considered the transaction graph, \eg in the quantitative analysis due to Ron and Shamir~\cite{2013_ron_transaction_analysis} or the assessment of payment anonymity by Meiklejohn \etal~\cite{2013_meiklejohn_anonymity_analysis}.
Several works surveyed scalability limitations and corresponding remedies~\cite{2016_tschorsch_beyond,2019_monrat_scalability,2020_zhou_scalability,2020_hafid_scalability}.
Higher-level data semantics were the subject of further analyses.
After Shirriff~\cite{2014_shirriff_blogpost} highlighted the presence of non-financial blockchain content hidden within transactions, Matzutt \etal~\cite{2016_matzutt_poster,2018_matzutt_contents} further formalized this aspect with a quantitative content analysis with a focus on potentially objectionable content on Bitcoin's blockchain.
Sward \etal~\cite{2017_sward_contents} concurrently investigated more sophisticated content insertion methods.
Meanwhile, Bartoletti \etal~\cite{2017_bartoletti_opreturn,2019_bartoletti_metadata} had investigated the constructive utilization of \code{OP\_RETURN} data.
Delgado-Segura \etal~\cite{2018_delgado_segura_utxo} and P{\'e}rez-Sol{\`a} \etal~\cite{2019_perez_sola_utxo} have investigated the UTXO sets of Bitcoin and related cryptocurrencies.
Finally, Lu \etal~\cite{2020_lu_generic_superlight} applied game theory to investigate the potential for even lighter clients.

\textbf{Lightweight Payment Schemes.}
To scale to large transaction throughputs, one branch of related work aims to avoid on-chain transactions to the largest extent possible without losing the trustworthiness of payments.
Poon and Dryja presented the Bitcoin Lightning Network~\cite{2016_poon_lightning} for this purpose, which since has sparked several further works~\cite{2015_decker_micropayment,2017_green_bolt,2017_khalil_payment_networks,2017_malavolta_payment_networks,2019_miller_payment_networks}.

\textbf{Countering Illicit Content.}
Matzutt \etal~\cite{2018_matzutt_thwarting} have assessed how to hinder the insertion of blockchain content in Bitcoin with only minimal changes to the consensus rules.
Most related countermeasures deal with illicit content after it has been inserted into the blockchain.
Florian \etal~\cite{2019_florian_erase} propose that nodes locally remove content they deem objectionable and investigate how to outsource the validation of removed transactions to other nodes.
A different branch of research revolves around redactable blockchains, which allow to remove transactions either based on the decision of one or few redactors~\cite{2017_ateniese_redactable_blockchain}, removal votings among miners~\cite{2019_deuber_redactable_blockchain}, or based on redaction policies specified by the transaction owner herself~\cite{2019_dorri_mofbc,2019_derler_chameleon_abe}, \eg to deal with privacy concerns.

\textbf{Blockchain Sharding.}
One special measure to increase the scalability of blockchain systems is \emph{sharding}.
In traditional blockchain systems, all nodes redundantly replay all events locally.
Hence, they do not have to especially trust other nodes at the cost of limited scalability.
Blockchain sharding, contrarily, partitions the control over the blockchain over time.
Bitcoin-NG~\cite{2016_eyal_bitcoinng} initially proposed to repurpose the mining process such that a successful miner becomes a leader capable of subsequently issuing blocks at a much higher frequency until the mining process yields the next leader.
Following this proposal, Luu \etal~\cite{2016_luu_elastico} proposed to use entropy periodically derived from blockchain data to randomly elect small committees that are subsequently responsible for processing a subset of transactions on behalf of the whole network.
Further work, such as OmniLedger~\cite{2018_kokoris_omniledger} and RapidChain~\cite{2018_zamani_rapidchain}, has improved upon this initial design since.

\textbf{Takeaway.}
Lightweight payment schemes provide an orthogonal approach to tackle blockchain scalability by avoiding on-chain transactions.
Further research also investigates illicit content on blockchains.
Corresponding countermeasures, except for the local erasure proposed by Florian \etal~\cite{2019_florian_erase}, usually neglect the possibility of objectionable content also entering the UTXO set, \ie the state maintained even when pruning a blockchain.
Finally, the epochs typically used in sharding blockchains have proven themselves a valuable building block for synchronizing multiple nodes and can be reused, \eg for coordinating periodically recurring tasks.

\section{Requirements for a Secure and Retrofittable Block-Pruning Scheme}%
\label{sec:requirements}

Blockchain systems require that sufficiently many nodes maintain a \emph{full local blockchain copy} (\cf~Section~\ref{sec:background}).
While this initial design becomes massively burdening for these nodes as well as joining nodes, multiple approaches to fully pruning obsolete data have been proposed (\cf~Section~\ref{sec:sota:proposals}).
However, none of these approaches can be adapted to directly provide similar optimizations for established systems (\eg, Bitcoin) without provoking major incompatibilities.
Furthermore, to the best of our knowledge no current pruning scheme considers \emph{data semantics}, \eg UTXO set pollution or higher-level application data.
To realize \emph{fully compatible} extensions for the network-wide pruning of obsolete data in existing cryptocurrencies while \emph{maintaining already established security levels} and enabling \emph{additional protection}, we identify the following requirements and design goals:

\newcounter{DesignGoals}
\refstepcounter{DesignGoals}\label{goal:scalability}
\newcommand{\goalscalability}{\textbf{(G\ref{goal:scalability})}\xspace}
\refstepcounter{DesignGoals}\label{goal:correctness}
\newcommand{\goalcorrectness}{\textbf{(G\ref{goal:correctness})}\xspace}
\refstepcounter{DesignGoals}\label{goal:verifiability}
\newcommand{\goalverifiability}{\textbf{(G\ref{goal:verifiability})}\xspace}
\refstepcounter{DesignGoals}\label{goal:compatibility}
\newcommand{\goalcompatibility}{\textbf{(G\ref{goal:compatibility})}\xspace}
\refstepcounter{DesignGoals}\label{goal:semantics}
\newcommand{\goalsemantics}{\textbf{(G\ref{goal:semantics})}\xspace}

\textbf{\goalscalability Scalability.}
To be effective, pruning schemes must provide improvements for \emph{all} metrics discussed in Section~\ref{sec:motivation}, \ie storage and bandwidth demands for joining and block-serving nodes, processing costs, and synchronization time.

\textbf{\goalcorrectness Correctness.}
Starting from the genesis block, each joining node must obtain the same local state, with or without the block-pruning scheme enabled, to ensure that the network's consensus about accepted transactions is kept intact.
In particular, the node must learn about all accepted, non-obsolete events, and it must not accept any false events.

\textbf{\goalverifiability Verifiability.}
As security is a top priority, pruning schemes must keep joining nodes capable of verifying the correctness of the synchronization process even in the presence of adversaries.
Here, we require that block-pruning schemes do not reduce the security of the overall blockchain system. 

\textbf{\goalcompatibility Compatibility.}
Popular and long-living blockchain systems are especially affected by scalability limitations.
Instead of proposing new systems (\cf~Section~\ref{sec:sota:proposals}), all changes should be applicable to existing blockchains, especially Bitcoin, even during operation.
Preferably, the scheme is opt-in, \eg as achievable via velvet forks (\cf~Section~\ref{sec:background}).

\textbf{\goalsemantics Data Semantics.}
Block-pruning schemes may assume more tasks and responsibilities than simply pruning obsolete data.
One crucial aspect is the handling of non-financial blockchain data.
Block-pruning schemes should be aware of the semantics of such data, \ie mitigate negative impacts of UTXO set pollution where possible and avoid impeding higher-level applications that make use of (prunable) \code{OP\_RETURN} transaction outputs.

\section{\name Overview}%
\label{sec:overview}

\begin{figure}[t]
    \centering
    \includegraphics{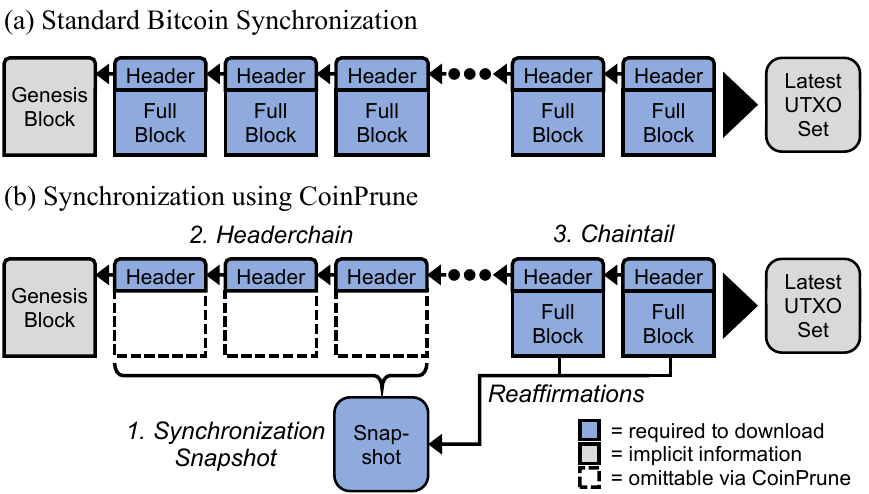}
    \caption{
        High-level design overview of \name.
        Instead of downloading and verifying all full blocks, joining nodes obtain a recent snapshot in a trustworthy manner due to its on-chain reaffirmations by multiple miners.
    }
    \label{fig:overview}
    \vspace{-1.5em}
\end{figure}

To address the challenges resulting from the ever-increasing size of existing blockchains, we present \emph{\name}, our secure, snapshot-based block-pruning scheme that is gradually deployable without protocol-breaking changes, \eg to Bitcoin.
In this section, we overview \name's design and, hereafter, we present its block-pruning scheme and the way we handle additional data semantics in the following sections.

\name is designed to transfer scalability improvements of novel blockchain designs (\cf~Section~\ref{sec:sota:proposals}) to Bitcoin while keeping compatibility a priority.
We now describe how \emph{\name nodes}, \ie Bitcoin nodes that additionally support \name, jointly maintain recent snapshots on the blockchain and how new nodes can bootstrap securely via those snapshots.
Finally, we outline benefits for the whole network.

\textbf{Snapshot Maintenance.}
\name nodes periodically create \emph{snapshots} of their UTXO set.
These snapshots are served to joining nodes instead of the full blockchain history for reduced storage, bandwidth, and processing requirements~\goalscalability.
They are tied to the current \emph{block height}, \ie the position of the most recent block in the blockchain, and contain a well-ordered UTXO set for synchronization~\goalcorrectness and verification~\goalverifiability purposes, respectively.
To prevent malicious nodes from distributing invalid snapshots, \eg in an attempt to multiply their funds, \name requires snapshots to be \emph{publicly announced} to the blockchain by referencing a cryptographic \emph{identifier} of each snapshot on-chain.
\name miners place these announcements in their blocks' \emph{coinbase transactions}, which the miners issue to mint new coins.
By utilizing an existing field within the coinbase transactions, which may contain \SI{100}{\byte} of arbitrary data, we keep \name Bitcoin-compatible~\goalcompatibility.
Other \name miners independently do the same, which causes nodes deriving snapshots from the same UTXO set to \emph{mutually reaffirm} that snapshot's validity.
This approach creates positive-only feedback, \ie wrong snapshots are not rejected but tolerated and outpaced by valid snapshots' reaffirmations given an honest majority of \name miners.

\textbf{Bootstrapping Nodes.}
Instead of downloading all blockchain data, a joining node can now securely bootstrap in three steps, as shown in Figure~\ref{fig:overview}:
First, the node obtains a \emph{recent snapshot} either from its neighbors or through a snapshot-offering third party (\cf~Section~\ref{sec:security}).
For P2P-based snapshot acquisition, the node downloads the snapshot advocated by the majority of its neighbors.
Second, the node obtains the \emph{headerchain}, \ie the interconnected and lightweight block headers, to learn about the blockchain branch with the most PoW in it.
Third, the node downloads the \emph{chaintail}, \ie the full blocks following the snapshot's block height.
Via the chaintail, the joining node can
\begin{enumerate*}
    \item{catch up with recent transactions and}
    \item{inspect the full blocks for snapshot reaffirmations.}
\end{enumerate*}
If the joining node observes sufficiently many reaffirmations of the obtained snapshot, it accepts this snapshot and concludes the initial synchronization process.
Otherwise, the node discards the insecure snapshot and retries while reconnecting to new neighbors.

\textbf{Global Block Pruning.}
Since joining nodes can securely bootstrap from the headerchain, the snapshot, and the chaintail, \emph{all} \name nodes may now \emph{safely prune older blocks} from before the snapshot.
As new snapshots are reaffirmed periodically, nodes may also prune aging snapshots as well without hurting the network health.
Single \emph{archival nodes} may still keep a full blockchain copy to retain the full and reliable queryability of also older data.

\textbf{Additional Data Semantics.}
\name is aware that
\begin{enumerate*}
    \item entries in the UTXO set may have been manipulated to cause UTXOs to be unspendable or even to hold illicit content and that
    \item small application-level data chunks stored in \code{OP\_RETURN} transactions are desirable to preserve but inherently prunable.
\end{enumerate*}
To account for these higher-level data semantics~\goalsemantics, \name nodes can \emph{locally obfuscate} most manipulable identifiers without impeding the nodes' capability of validating pending transactions and they maintain an additional \emph{application data storage}, which contains all \code{OP\_RETURN} transaction outputs and their context (\cf~Section~\ref{sec:data-semantics}).

\vspace{-0.5em}
\section{Retrofittable Block Pruning}%
\label{sec:pruning}

We now present \name's block-pruning scheme, which is designed for a retrospective deployment to existing cryptocurrencies.
We discuss the data management of \name nodes, how these nodes coordinate the pruning process, and how new nodes can now bootstrap efficiently.

\vspace{-0.5em}
\subsection{Adapted Data Management}%
\label{sec:pruning:data}

To understand \name in detail, we discuss the layout of its snapshots and the required changes to the nodes' local data management stemming from the pruning of older blocks.

\textbf{Snapshot Creation.}
Each snapshot corresponds to a specific block height, meaning that it represents a serialization of the UTXO set obtained from processing all blocks up to and including that height.
A \name snapshot consists of a simple header and multiple chunks of serialized UTXOs, and it is referenced on-chain by a cryptographic identifier.
The header holds the snapshot's corresponding block height, that block's identifier, and the number of chunks in the snapshot.
The identifier is a special hash value created over the snapshot's header and chunks to uniquely represent the snapshot in a succinct manner.
First, the header and each chunk are hashed individually using Bitcoin's HASH256 function (SHA256 applied twice).
Then, the snapshot identifier is the hash value of the concatenation of these hash values.
Using this simple snapshot serialization, joining nodes are immediately aware of all available chunks and can independently request individual chunks from different neighbors in parallel.
Further, we limit chunk sizes to \SI{1}{\mega\byte} akin to Bitcoin's maximum block size.

\textbf{Persisted Information.}
By shifting to a snapshot-based synchronization process, nodes may now prune older data by deleting the full blocks prior to the snapshot's block height.
However, the nodes must remain capable of serving the full headerchain to joining nodes.
Before pruning blocks, these nodes thus need to persist some information currently held by Bitcoin's block index.
These are block identifiers, headers, block heights, the amount of PoW, the number of transactions, and the block's timestamp.
Persisting this data, a recent snapshot, and the chaintail of not-yet-prunable full blocks is now sufficient to securely bootstrap joining nodes.

\begin{figure}[t]
    \centering
    \includegraphics{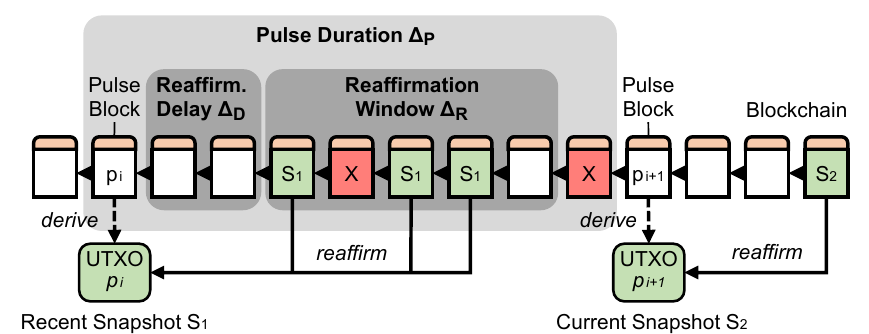}
    \vspace{-1.8em}
    \caption{
        Our pulse-based coordination triggers the creation of new snapshots.
        Any invalid or delayed reaffirmations are ignored.
    }
    \label{fig:pruning:slotting}
    \vspace{-1.6em}
\end{figure}

\vspace{-0.6em}
\subsection{Coordination of \name Nodes}%
\label{sec:pruning:coordination}

\name establishes trust in recent snapshots by having miners mutually reaffirm these snapshots' correctness.
Joining nodes then use these reaffirmations as their trust anchor when deciding whether to synchronize based on a particular snapshot.
As shown in Figure~\ref{fig:pruning:slotting}, \name miners coordinate based on \emph{pulse blocks $p_i$}, which are issued in constant intervals $\Delta_P$, \eg every \num{10000} blocks.
These pulse blocks trigger the creation of a new snapshot and its subsequent mutual reaffirmation at fixed positions on the blockchain, \ie all nodes act based on the same information.

Each snapshot is attached to a pulse block, \ie \name nodes subsequently reaffirm the snapshot derived from the pulse block's corresponding UTXO set.
All \name miners reaffirm the last pulse block's attached snapshot by adding the snapshot's identifier to their blocks' coinbase fields during a relatively short \emph{reaffirmation window $\Delta_R$}.
During $\Delta_R$, multiple concurring snapshot identifiers might occur if an adversary attempts to get an invalid snapshot reaffirmed.
However, we assume an honest majority among \name miners, just like in the overall Bitcoin network, and thus the genuine snapshot is expected to accumulate reaffirmations the fastest.
We further enforce an \emph{acceptance threshold $k$}, \ie a joining node only accepts the most reaffirmed snapshot if it was reaffirmed at least $k$ times during $\Delta_R$.
If no snapshot reaches $k$ reaffirmations during $\Delta_R$, this pulse is invalid, and pruning is delayed until the next pulse starts.
The goals behind these measures are 
\begin{enumerate*}
    \item preventing a dishonest minority from outpacing an honest majority during snapshot reaffirmation and
    \item preventing a dishonest majority stemming from low overall \name support from successfully reaffirming any snapshot during $\Delta_R$ (\cf~Section~\ref{sec:security}).
\end{enumerate*}
Finally, the reaffirmation window may be delayed by a small \emph{reaffirmation delay $\Delta_D$}, \eg $\Delta_D\!=\!\SI{6}{blocks}$, to ensure that accidental forks affecting the pulse block $p_i$ are resolved before creating a snapshot.

\vspace{-0.6em}
\subsection{Bootstrapping New Nodes}%
\label{sec:pruning:bootstrapping}

The reaffirmations periodically published on Bitcoin's blockchain allow joining nodes to bootstrap as follows.
First, the node obtains a recent snapshot.
The node can now securely acquire a recent snapshot from its neighbors using off-chain P2P requests or through external means, \eg mirror servers.
Second, the joining node downloads and verifies the headerchain from its neighbors to learn about the blockchain branch with the most PoW in it, as is already done by Bitcoin~\cite{2015_bitcoin_v0_10_synchronization}.
Third, instead of downloading and processing all blockchain data, the joining node applies the previously obtained snapshot in good faith to initially fill its UTXO set.
Finally, the joining node fetches and processes the \emph{chaintail}, \ie the remaining full blocks succeeding the snapshot's block height, to finalize synchronizing its UTXO set.
During this full synchronization phase, the joining node additionally inspects the chaintail's coinbase transactions for reaffirmations of its applied snapshot.
If the node learns that its snapshot was the most-reaffirmed one during $\Delta_R$ and that it was reaffirmed at least $k$ times, it accepts the snapshot and concludes the bootstrapping step.
Otherwise, the joining node aborts and obtains a different snapshot from another source, \eg by connecting to a new set of neighbors.

\vspace{-0.5em}
\section{Handling Application-Level Data}%
\label{sec:data-semantics}

We now present \name extensions for handling non-financial data.
We first show how local \emph{obfuscation} effectively removes illicit contents from the UTXO set without losing Bitcoin compatibility.
Then, we discuss how \name preserves intended application-level data, \eg \code{OP\_RETURN} data.

\vspace{-0.6em}
\subsection{Obfuscation of Illicit Data from the UTXO Set}%
\label{sec:data-semantics:obfuscate}

As shown by Florian \etal~\cite{2019_florian_erase}, nodes can locally delete objectionable content from their copy of the blockchain and their UTXO set at the cost of outsourcing the validation of corresponding transactions to other nodes.
We extend upon this idea and integrate a simple \emph{UTXO obfuscation scheme} into \name that prevents nodes from recovering any objectionable data from snapshots while remaining capable of using obfuscated UTXOs during transaction validation.

\begin{figure}[t]
    \centering
    \includegraphics{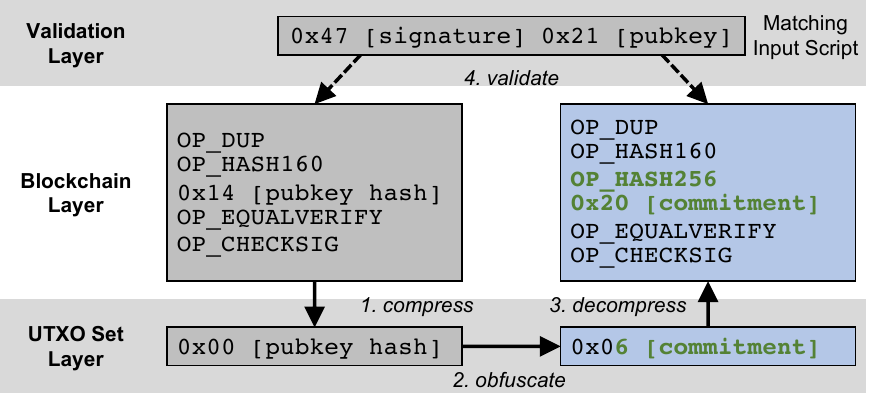}
    \vspace{-1.8em}
    \caption{
        \name obfuscates hash-based, compressed UTXOs (here P2PKH) by additionally applying \texttt{OP\_HASH256} to the script's mutable value.
        Nodes account for this change during decompression and can still validate the expected input script spending the obfuscated UTXO without further alterations.
    }
    \label{fig:data-semantics:example}
    \vspace{-1.5em}
\end{figure}

\textbf{Construction.}
In previous work, we proposed \emph{self-verifying blockchain identifiers}~\cite{2018_matzutt_thwarting} to harden the mutable fields of a transaction output script against easy content insertion.
Essentially, this measure forces transaction creators to only release \emph{one-way commitments} to these mutable values while allowing full nodes to detect further tampering easily.
As indicated by Florian \etal~\cite{2019_florian_erase}, \name can make use of similar and simple local UTXO set obfuscation, which extends to the snapshots sent to joining nodes:
Hash-based transaction outputs, such as P2PKH (\cf~Section~\ref{sec:background}), already commit to a later disclosed public key by recording only its HASH160 value on-chain, \ie the spender needs to open the commitment as part of the spending condition.
However, because this on-chain commitment remains mutable and has been used for content insertion~\cite{2018_matzutt_contents}, we \emph{locally commit} to these values again \emph{before} including the results in the UTXO set.

More specifically, \name locally obfuscates mutable and potentially content-holding blockchain identifiers, as shown in Figure~\ref{fig:data-semantics:example}.
Before updating the UTXO set, each Bitcoin node already compresses any transaction output corresponding to one of the six predefined patterns (\cf~Section~\ref{sec:background}) and uses one byte to denote the compression case applied, \ie the values \code{0x00} to \code{0x05} (Step~1).
Otherwise, the node keeps the uncompressed script and prepends it with its length plus six (to account for the six compression cases).
Using our obfuscation scheme (Step~2), \name-supporting nodes now additionally apply Bitcoin's HASH256 function to the respective mutable values in any standard-compliant, hash-based transaction output.
Consequently, the node cannot restore the value due to the pre-image resistance of SHA256.
Besides P2PKH, \name can thus also obfuscate pay-to-script-hash (P2SH) scripts, which commit to an entire script defining a more flexible spending condition, and their segregated witness-based counterparts P2WPKH and P2WSH~\cite{2016_lombrozo_segwit} this way.
Combined, these script types account for \btcstatEvalStorageAntiContentsFractionObfuscated of all current UTXOs (\cf~Section~\ref{sec:eval:pruning}).
We account for the proper decompression of obfuscated values to maintain compatibility with Bitcoin by introducing four new compression cases, \eg an obfuscated P2PKH value will be denoted by \code{0x06} instead of \code{0x00} (\cf~Figure~\ref{fig:data-semantics:example}).
During decompression (Step~3), the node can now detect obfuscated values and add the required \code{OP\_HASH256} operation as the second local commitment layer.
This way, the node remains capable of validating incoming input scripts even if the spender is unaware of the obfuscation.

\textbf{Limitations.}
Our obfuscation scheme relies on the fact that most UTXOs already contain a commitment instead of, \eg raw public keys, \ie the transaction input spending a UTXO will open the commitment.
In this case, we can indeed add another commitment layer without further changes.
However, albeit deprecated, Bitcoin still permits P2PK and P2MS transaction outputs, which contain one or multiple raw public keys, respectively~\cite{2014_okupski_bitcoin_reference}.
Since the main goal of \name is remaining retrofittable to Bitcoin, we cannot adapt the required transaction input scripts.
Furthermore, we refrain from caching raw public keys before committing to them in the UTXO set since these public keys can also be manipulated~\cite{2018_matzutt_contents}.
Hence, our scheme cannot support the obfuscation of these or any non-standard transaction outputs without losing its deployability as a velvet fork.
Still, our obfuscation scheme already covers \btcstatEvalStorageAntiContentsFractionObfuscated of the current UTXOs (\cf~Section~\ref{sec:eval:pruning}).

\textbf{Future Potentials.}
In case of widespread adoption of \name, it may become possible to shift away from ensuring deployability as a velvet fork and propose further changes affecting the UTXO set.
First, strictly enforcing transaction standardness and rejecting legacy P2PK and P2MS transaction outputs can improve the coverage of our obfuscation scheme.
Second, currently non-obfuscatable UTXOs already contained in the UTXO set could be rewritten to become obfuscatable.
This rewriting is currently not possible since legacy nodes would not be aware that they have to create a different script when attempting to spend their old coins.
Finally, if the network agrees on carefully rewriting the UTXO set to a certain extent, future work can investigate further means to counteract UTXO set pollution.
For instance, nodes could completely drop UTXOs flagged for encoding illicit content or repossess old entries that only hold negligible values and thereby bloat the UTXO set.
However, such modifications would have strong implications for Bitcoin in general, demanding a comprehensible yet minimal rule set for allowed modifications approved by a large majority of nodes.

\vspace{-0.5em}
\subsection{Preservation of Application-Level Data}%
\label{sec:data-semantics:preserve}

Bitcoin explicitly enables the insertion of small, application-specific data chunks in a transaction using \code{OP\_RETURN} transaction outputs (\cf~Section~\ref{sec:background}).
Such transaction outputs are inherently unspendable, and thus they are not included in the UTXO set.
Consequently, block-pruning schemes, such as \name, would remove this data and thereby break higher-level applications, \eg, audit systems~\cite{proof_of_existence,namecoin,2017_henze_dcam}.

Thus, a second \name extension provides an \emph{application data storage} that realizes a lookup table of past \code{OP\_RETURN} transaction outputs with their context (block and transaction).
Each entry in this lookup table consists of
\begin{enumerate*}
    \item the \code{OP\_RETURN} payload as well as the identifiers of the corresponding
    \item transaction and
    \item block.
\end{enumerate*}
As nodes still maintain the headerchain, this information suffices to associate application data with a specific transaction and its inclusion time.

Applying this extension changes the semantics of a \name reaffirmation.
The application data storage is maintained in \SI{1}{\mega\byte}-chunks just as snapshots are (\cf~Section~\ref{sec:pruning:data}), and thus nodes obtain its identifier in the same way.
However, the reaffirmation tags written to the blockchain are now derived by applying HASH256 to the concatenated snapshot identifier \emph{and} the application data storage's identifier.

\vspace{-0.5em}
\section{Seamless Integration into Bitcoin}%
\label{sec:integration}

\name's main feature is its immediate applicability to Bitcoin~\goalcompatibility.
In this section, we present our means to achieve \emph{gradual opt-in deployability} to Bitcoin via a velvet fork, assuming that a sufficient share of honest miners makes a rational choice to support \name, \eg to preserve storage.

\textbf{On-Chain Data.}
Although snapshot reaffirmations must be publicly announced on Bitcoin's blockchain, \name's utilization of only a block's coinbase field prevents any protocol-breaking changes.
Full nodes that are unaware of \name will ignore any snapshot reaffirmation, and \name nodes will never reject blocks containing incorrect reaffirmations.
Instead, they will try to outpace incorrect reaffirmations with legitimate ones.
Hence, our scheme fulfills the requirements for a gradually deployable velvet fork~\cite{2017_kiayias_velvet_forks,2018_zamyatin_velvet_forks}.
To prevent \name nodes from confusing snapshot reaffirmations with other coinbase data, we propose to encapsulate the reaffirmation accordingly, \eg using a unique prefix and separators such as \code{CoinPrune/[reaffirmation]/}.
While \name has to potentially share the coinbase field with other applications, \eg merged mining~\cite{2017_judmayer_merged_mining}, its messages remain distinguishable from messages of other applications.

\textbf{Peer-to-Peer Protocol.}
Even though \name allows for external snapshot sources, most joining nodes will likely rely on Bitcoin's network to obtain their initial snapshot.
To enable this Bitcoin-intrinsic snapshot acquisition, we extend Bitcoin's P2P protocol~\cite{2017_bitcoin_p2p} with an additional \code{GETSTATE} message type sent by \name nodes.
Joining nodes send a \code{GETSTATE} message to each neighbor to learn about available recent snapshots.
Each neighbor responds with an inventory (\code{INV} message) that contains the hash values of the snapshot header and the chunks of their most recent available and successfully reaffirmed snapshot as \code{STATE} objects.
The joining node uses these \code{INV} messages to determine which snapshot to obtain and derives that snapshot's identifier.
Then, the node requests individual chunks of the most-advertised snapshot from its neighbors using sequences of \code{GETDATA} messages.
Finally, the node applies the snapshot once all chunks are available.
For increased compatibility, we restrict chunk sizes to \SI{1}{\mega\byte}, \ie Bitcoin's old maximum block size, and we introduce a new service flag for Bitcoin's \code{VERSION} handshake to avoid sending unknown messages to \name-unaware nodes.

\vspace{-0.7em}
\section{Security Discussion}%
\label{sec:security}

In this section, we discuss how \name helps nodes bootstrap correctly and securely~\goalcorrectness based on verifiable snapshots~\goalverifiability, which are maintained by an honest majority of \name miners.
We first show that the on-chain reaffirmations created by our scheme reliably reference snapshots from arbitrary sources to be used for bootstrapping joining nodes more efficiently.
Second, we discuss the influence of the positive-only feedback created by \name miners mutually reaffirming the current snapshot has on its trustworthiness.
We extend our previous work~\cite{2020_matzutt_coinprune} by an empirical simulation, showing that \name can achieve similar security properties as Bitcoin if at least \SI{32}{\percent} of Bitcoin nodes support the scheme.
Finally, we discuss the P2P-level security of \name.

\begin{figure}[t]
    \centering
    \includegraphics{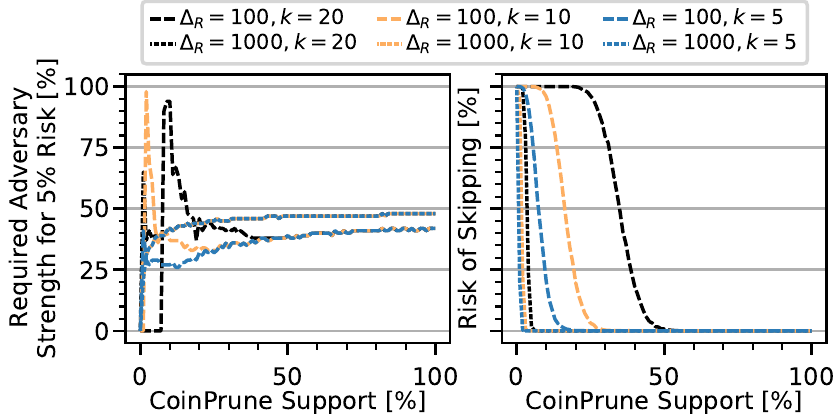}
    \vspace{-.8em}
    \caption{%
        Larger reaffirmation windows increase \name's resilience against adversaries independent from the acceptance threshold.
    }
    \label{fig:eval:security}
    \vspace{-1.7em}
\end{figure}

\textbf{Verifying Snapshot Validity.}
\name{}'s on-chain reaffirmations enable joining nodes to reliably verify a snapshot's correctness.
To this end, \name uses a cryptographic hash function in a layered fashion to obtain snapshot identifiers, thereby hashing each snapshot chunk individually.
This approach enables joining nodes to notice any manipulation of individual chunks, \eg performed by malicious nodes, during their initial synchronization.
As the snapshot identifier also covers the snapshot's metadata and the hash values of all chunks, joining nodes can further derive at which block height the snapshot was created and verify that they obtained all original chunks.
Thus, adversaries cannot deceive joining nodes to accept invalid or inconsistent snapshots, \eg to reinstate spent UTXOs.
Consequently, nodes can safely apply valid snapshots regardless of their source, \ie the Bitcoin network or a third party, such as a snapshot-mirroring website.

\textbf{Adversary Resilience.}
The reaffirmation tags \name miners add to their blocks allow joining nodes to verify that an obtained snapshot has indeed been reaffirmed.
However, an adversary may attempt to reaffirm an invalid snapshot.
In Section~\ref{sec:pruning:coordination}, we introduced the acceptance threshold $k$ and the reaffirmation window $\Delta_R$ to prevent an adversary from successfully reaffirming an invalid snapshot.
To investigate the effectiveness of these means, we simulated a random mining process of \num{1000} active miners with an increasing support of \name among miners and an increasing adversarial influence among \name miners.

Namely, we consider a support of \name among miners of $\SI{0}{\percent} \le f_C \le \SI{100}{\percent}$ that grows in increments of $\SI{1}{\percent}$.
For each sampling point, we assume a growing fraction $f_A$ of these \name miners to be controlled by an adversary, where we again vary $\SI{0}{\percent} \le f_A \le \SI{100}{\percent}$ in increments of $\SI{1}{\percent}$.
For instance, if $f_C = \SI{25}{\percent}$ and $f_A = \SI{10}{\percent}$, then we assume that \num{250} out of \num{1000} active miners reaffirm a snapshot, but \num{25} of those miners will reaffirm an invalid one.
We repeated each simulation \num{1000} times and considered reaffirmation windows $\Delta_R \in \{100, 1000\}$ and acceptance thresholds $k \in \{5, 10, 20\}$, respectively, and investigated how often
\begin{enumerate*}
    \item a joining node would accept the correct snapshot,
    \item the adversary was able to successfully reaffirm an invalid snapshot, and
    \item no snapshot was reaffirmed successfully within the reaffirmation window, \ie the pulse is skipped without pruning.
\end{enumerate*}

In Figure~\ref{fig:eval:security}, we show the minimum fraction of adversaries $f_A$ required such that at least \SI{5}{\percent} of snapshots a joining node accepts are in fact reaffirmed by the adversary (left-hand side) and the worst-case risk over all $f_A$ that no snapshot was reaffirmed successfully (right-hand side).
Our results show that a longer reaffirmation window of $\Delta_R\!=\!1000$, \ie roughly one week, raises the required $f_A$ for compromising at least \SI{5}{\percent} of reaffirmed snapshots to $f_A\!\ge\!\SI{46}{\percent}$ for $f_C\!\ge\!\SI{31}{\percent}$ with $f_A\!=\!\SI{48}{\percent}$ for full \name support independent of the acceptance threshold $k$.
For the smaller $\Delta_R$, increasing $k$ has the desired effect of raising the required $f_A$ in case of low \name adoption, but this comes at the cost of disproportionally impeding the operation of \name due to a large number of skipped pulses.
We can counteract this effect by decreasing the pulse duration $\Delta_P$, and thus a \emph{dynamic} approach adapting $\Delta_R$ and $\Delta_P$ based on a previously sampled support level $f_C$ is promising to combine the positive effects of short and long reaffirmation windows.
During phases of low support, \eg $f_C\!<\!\SI{10}{\percent}$, we can operate with $\Delta_R\!=\!100$, but aggressively retry to reaffirm snapshots by also setting $\Delta_P\!=\!\Delta_R\!=\!100$.
If $f_C$ was sufficiently large during the last pulse, we can relax \name's aggressiveness and set, \eg $\Delta_R\!=\!\num{1000}$ and $\Delta_P\!=\!\num{10000}$.
Finally, a voting phase among miners, similar to that preceding the rollout of P2SH support~\cite{2012_andresen_ptsh}, can ensure a sufficient initial adoption rate $f_C$.

\textbf{P2P Attacks.}
Nodes announce their support of \name during the version handshake (\cf~Section~\ref{sec:integration}) to integrate well with Bitcoin's P2P protocol~\goalcompatibility.
Our scheme further adds only two message types not understood by vanilla Bitcoin nodes, which are only exchanged with \name nodes.
First, joining nodes send an additional \code{GETSTATE} message.
Second, contacted nodes answer with an \code{INV} message containing the newly introduced \code{STATE} objects.
Both message types are in line with the design of Bitcoin's P2P protocol, and considerations regarding its DoS resilience or Eclipse attacks~\cite{2015_heilman_eclipse_attacks} translate directly to \name.
Furthermore, as shown above, an adversary cannot undetectably advertise an invalid or partially manipulated snapshot.
Contrarily, the P2P layer provides an additional defense layer in the rare case where an adversary controlling only a minority of \name miners successfully reaffirms an invalid snapshot by chance.
In this case, honest nodes will not advertise the invalid snapshot in their \code{INV} message and thus provoke a pulse to be effectively skipped instead of spreading invalid snapshots.
Whenever a joining node suspects an attack, it will start over and bootstrap with a new neighbor set.
When reconnecting one of our commodity PCs (\cf~Section~\ref{sec:eval:setup}) to the Bitcoin network every \SI{30}{\minute} between Jan 26 and Feb 1, 2021 it took the client $\sim$\SI{68}{\second} on average to connect to eight new neighbors.
While the client continually establishes new connections afterward, it establishes its tenth outgoing connection after only $\sim$\SI{154}{\second} on average, which shows the feasibility of our approach even when accounting for early disconnects.

\textbf{Takeaway.}
Through \name, joining nodes can securely obtain the snapshots for initial synchronization from arbitrary sources given their on-chain reaffirmations' reliability.

\begin{figure*}[t]
    \begin{minipage}[t]{0.32\textwidth}
        \centering
        \includegraphics{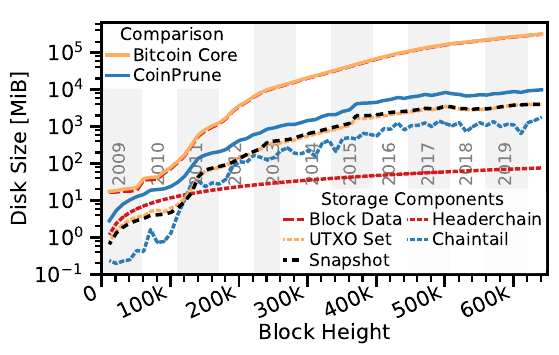}
        \vspace{-1.2em}
        \caption{%
            Currently, our scheme already reduces storage requirements by two orders of magnitude.
        }
        \label{fig:eval:disksize}
        \vspace{-1.5em}
    \end{minipage}
    \hfill
    \begin{minipage}[t]{0.32\textwidth}
        \centering
        \includegraphics{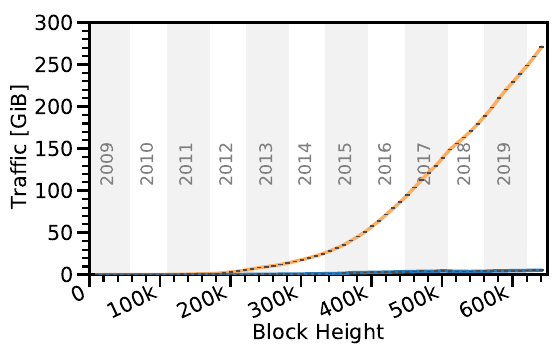}
        \vspace{-1.2em}
        \caption{%
            \name allows bootstrapping with vastly reduced amounts of traffic, unburdening all nodes.
        }
        \label{fig:eval:traffic}
        \vspace{-1.5em}
    \end{minipage}
    \hfill
    \begin{minipage}[t]{0.32\textwidth}
        \centering
        \includegraphics{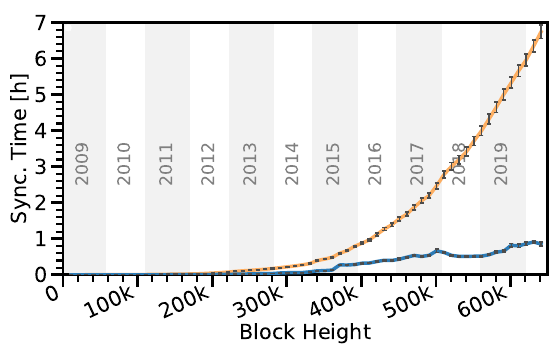}
        \vspace{-1.2em}
        \caption{%
			Our scheme reduces initial synchronization times, yet the snapshot size impacts performance.
        }
        \label{fig:eval:sync-time}
        \vspace{-1.5em}
    \end{minipage}
\end{figure*}

\vspace{-0.5em}
\section{Performance Evaluation}%
\label{sec:eval}

We now demonstrate that \name enables massive performance savings for Bitcoin nodes~\goalscalability.
After describing our testbed setup, we present the storage savings achieved for all nodes.
Further, we show that traffic and synchronization times for joining nodes are massively reduced as well.

\vspace{-0.5em}
\subsection{Testbed Setup for Synchronization Measurements}%
\label{sec:eval:setup}

We use our proof-of-concept implementation of \name based on Bitcoin Core v0.17.1 to run measurements on a server (2$\times$ Intel Xeon E5-2630 v4, \SI{32}{\giga\byte} RAM, \SI{8}{\tera\byte} Seagate IronWolf ST8000VN0022-2EL112), which bootstraps from eight identical commodity PCs (Intel Core2 Quad Q9400, \SI{8}{\giga\byte} RAM, \SI{500}{\giga\byte} Hitachi Deskstar 7K500) running \name as well.
All devices are interconnected via a Linksys SLM2024 Gigabit switch.
For our measurements, we consider snapshots being created every \num{10000} blocks up until a block height of \num{640000} (Jun~17, 2020) with \num{1000} additional blocks ($\sim$\SI{1}{week} of blocks) as our chaintail.
We perform the synchronization using vanilla Bitcoin Core in one go and synchronize via \name based on the aforementioned snapshots.
While storage requirements are fully determined by the blockchain data, synchronization times and traffic may vary.
Hence, the latter were averaged over five independent runs, and error bars denote the \SI{99}{\percent} confidence intervals.
We omitted to check the coinbase field for the snapshot identifier to be able to use Bitcoin's real blockchain for our measurements.

\vspace{-0.4em}
\subsection{Storage Savings}%
\label{sec:eval:pruning}

\name allows all Bitcoin nodes to prune older blocks in exchange for maintaining a recent snapshot and the headerchain to serve joining nodes.
In Figure~\ref{fig:eval:disksize}, we depict how the main contributors to Bitcoin's storage demand changed over time compared to the serialized snapshot and headerchain required to operate \name.
From this, we derive the overall storage requirements for Bitcoin Core and \name, respectively.
For Bitcoin's storage requirements, we consider the heavily dominating \code{blocks} folder containing raw block data, information required to rewind blocks efficiently, and the block index, as well as the \code{chainstate} folder holding the UTXO set.
In contrast to this, \name needs to store one serialized snapshot and the serialized headerchain, as well as the UTXO set and chaintail for live operation.
Our measurements show that the sizes of serialized snapshots align well with those of Bitcoin's UTXO set.
Minor variances stem from different encodings of both data structures.
Further, persisting the headerchain to reconstruct Bitcoin's block index comes at only negligible costs of \btcstatEvalStorageHeaderchainPerBlock per block, resulting in a headerchain size of \btcstatEvalStorageHeaderchainSize for our latest measurement.
Finally, considering block heights starting from \num{500000}, the chaintail has an average size of \btcstatEvalStorageMeanChaintailSizeLater.
Overall, \name nodes could thus historically reduce their storage requirements by \btcstatEvalStorageMeanReduction, with the largest absolute and relative savings, currently \btcstatEvalStorageMaxSave, at higher block heights.
These savings account for a decrease of two orders of magnitude, with the potential for becoming even larger as the blockchain grows.

\textbf{Overhead of Content Obfuscation.}
We consider our most recent snapshot (block height \num{640000}), which contains roughly \SI{66.2}{\million} UTXOs, and assess our obfuscation scheme's impact to protect against unwanted content~\goalsemantics.
Using our scheme, we can obfuscate \btcstatEvalStorageAntiContentsFractionObfuscated of all UTXOs, the vast majority of which hold P2PKH or P2PSH scripts.
However, the obfuscation of these scripts increases the snapshot size since we now store \SI{32}{\byte}-long commitments instead of the \SI{20}{\byte}-long values.
The considered snapshot grows from \btcstatEvalStorageAntiContentsSizeVanilla to \btcstatEvalStorageAntiContentsSizeObfuscated, \ie we inflict an overhead of \btcstatEvalStorageAntiContentsOverhead.
However, the overall storage savings and the protection against unwanted or even illegal blockchain content outweigh this overhead.

\textbf{Application Data Storage.}
To assess the growth of the application data storage~\goalsemantics, we serialized all \code{OP\_RETURN} transaction outputs up to the block height of our most recent snapshot according to our scheme presented in Section~\ref{sec:data-semantics:preserve}.
This serialized application data storage has a total size of \btcstatEvalStorageOpreturnSize and contains roughly \btcstatEvalStorageOpreturnNumber entries with an average size of \btcstatEvalStorageOpreturnAvgSize each.
Hence, by pruning old blocks while preserving application data specifically, \name continues to support higher-level applications at feasible costs.

\vspace{-.6em}
\subsection{Evaluation of Synchronization Performance}%
\label{sec:eval:performance}

Pruning obsolete data not only relieves Bitcoin nodes from storage depletion but also joining nodes benefit from a widespread adoption of \name.
As shown in Figure~\ref{fig:eval:traffic}, the reduced storage requirements directly translate to a reduction in traffic required to synchronize with the Bitcoin network.
For instance, synchronizing from a snapshot on block height \num{640000} with a chaintail length of \num{1000}, joining nodes only inflict \btcstatEvalTrafficBothMaxCompaction of traffic (dominated by the snapshot and the chaintail) when using \name, whereas legacy nodes would cause \btcstatEvalTrafficBothMaxVanilla of traffic to bootstrap successfully.
Over the whole blockchain, achievable savings average at \btcstatEvalTrafficMeanReduction.
Joining nodes currently have to obtain two orders of magnitude less data during initial synchronization using \name, which is largely dominated by acquiring the snapshot.

A similar trend can be observed for the overall synchronization time of joining nodes, \ie obtaining and verifying the headerchain, the snapshot, and the chaintail.
Figure~\ref{fig:eval:sync-time} shows that \name improves synchronization times over Bitcoin's whole history, resulting in savings of \btcstatEvalSynctimeMeanReduction on average for joining nodes.
Even though Bitcoin mitigates revalidating very old transactions due to its assumed-valid blocks (\cf~Section~\ref{sec:sota:updates}), joining nodes still must replay the whole transaction graph.
Contrarily, \name avoids this step as well due to its reliance on snapshots.
Consequently, \name currently enables joining nodes to catch up with the Bitcoin network in \btcstatEvalSynctimeMaxCompaction instead of \btcstatEvalSynctimeMaxVanilla using standard Bitcoin.
This time increases when a node joins toward the end of a pulse due to a longer chaintail.
However, Figure~\ref{fig:eval:sync-time} indicates that joining nodes benefit from a vastly improved initial synchronization performance even in this case.
This time saving is especially beneficial as initial synchronization is often considered a major scalability concern~\cite{2020_lopp_bitcoin_performance,2016_croman_blockchain_scalability}.

\textbf{Takeaway.}
The snapshot-based approach of \name unburdens both established and joining nodes from major overheads stemming from Bitcoin's bootstrapping process, including the required storage, traffic, and synchronization times.
Hence, \name establishes a secure and effective means to vastly improve Bitcoin's long-term durability.

\vspace{-0.5em}
\section{Conclusion}%
\label{sec:conclusion}

\name provides a gradually deployable and secure block-pruning scheme for established cryptocurrencies such as Bitcoin.
\name-supporting miners periodically create snapshots of Bitcoin's current UTXO set and mutually reaffirm the correct snapshot by cryptographically linking it to the blocks they mine.
Our scheme hinders an adversary's effort to distribute invalid snapshots by restricting reaffirmation activities to a short window after the snapshot creation and by requiring a minimum number of reaffirmations for joining nodes to accept a snapshot.
Our simulation shows that our scheme achieves security properties comparable to Bitcoin when $1/3$ of nodes in the Bitcoin network support \name.
Consequently, all \name-supporting nodes can completely prune obsolete data and reduce their required storage from currently \btcstatEvalStorageMaxVanillaHighlevel to \btcstatEvalStorageMaxCompactionHighlevel (a decrease of \SI{302}{\gibi\byte}), with potentially even larger saving potentials as the blockchain grows.
Due to the vastly reduced data to obtain and process, joining nodes can currently reduce their synchronization time by \btcstatEvalSynctimeMeanReductionHighlevel in our measurements.
Finally, \name handles non-financial data by
\begin{enumerate*}
    \item preserving application-level data from \code{OP\_RETURN} transaction outputs and
    \item obfuscating almost all objectionable and potentially illegal data that may still be present in the UTXO set after pruning.
\end{enumerate*}

\vspace{1.0em}
{
\footnotesize
\setstretch{.9}
\textsc{Acknowledgements.}
This work has been funded by the German Federal Ministry of Education and Research (BMBF) under funding reference numbers 16KIS0443, 16DHLQ013, and Z31 BMBF Digital Campus.
The funding under reference number Z31 BMBF Digital Campus has been provided by the German Academic Exchange Service (DAAD).
The responsibility for the content of this publication lies with the authors.\par
}

\vspace{-0.5em}
\bibliographystyle{IEEEtran}
\bibliography{paper}

\begin{IEEEbiography}[{\includegraphics[width=1in,height=1.25in,clip,keepaspectratio]{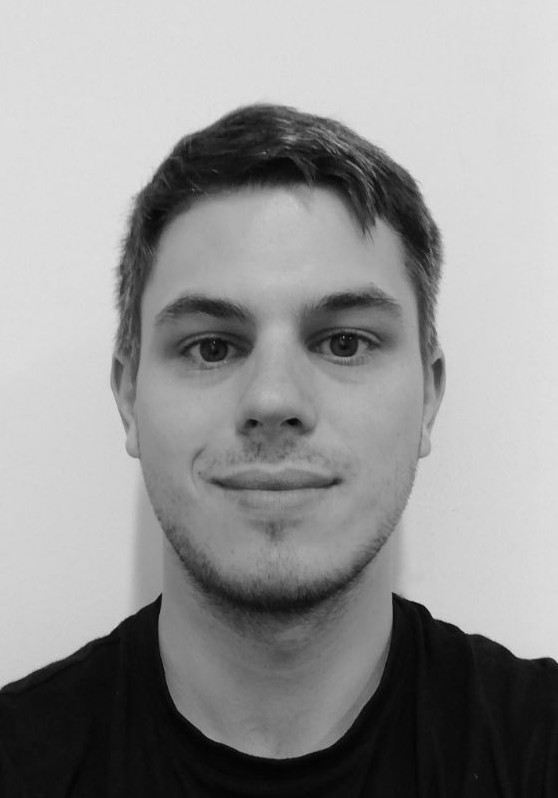}}]{Roman Matzutt}
received the B.Sc.\ and M.Sc.\ degrees in Computer Science from RWTH Aachen University.
He is a researcher at the Chair of Communication and Distributed Systems (COMSYS) at RWTH Aachen University.
His research focuses on the challenges and opportunities of accountable and distributed data ledgers, especially those based on blockchain technology, and means allowing users to express their individual privacy demands against Internet services.
\end{IEEEbiography}

\vspace{-1.5em}
\begin{IEEEbiography}[{\includegraphics[width=1in,height=1.25in,clip,keepaspectratio]{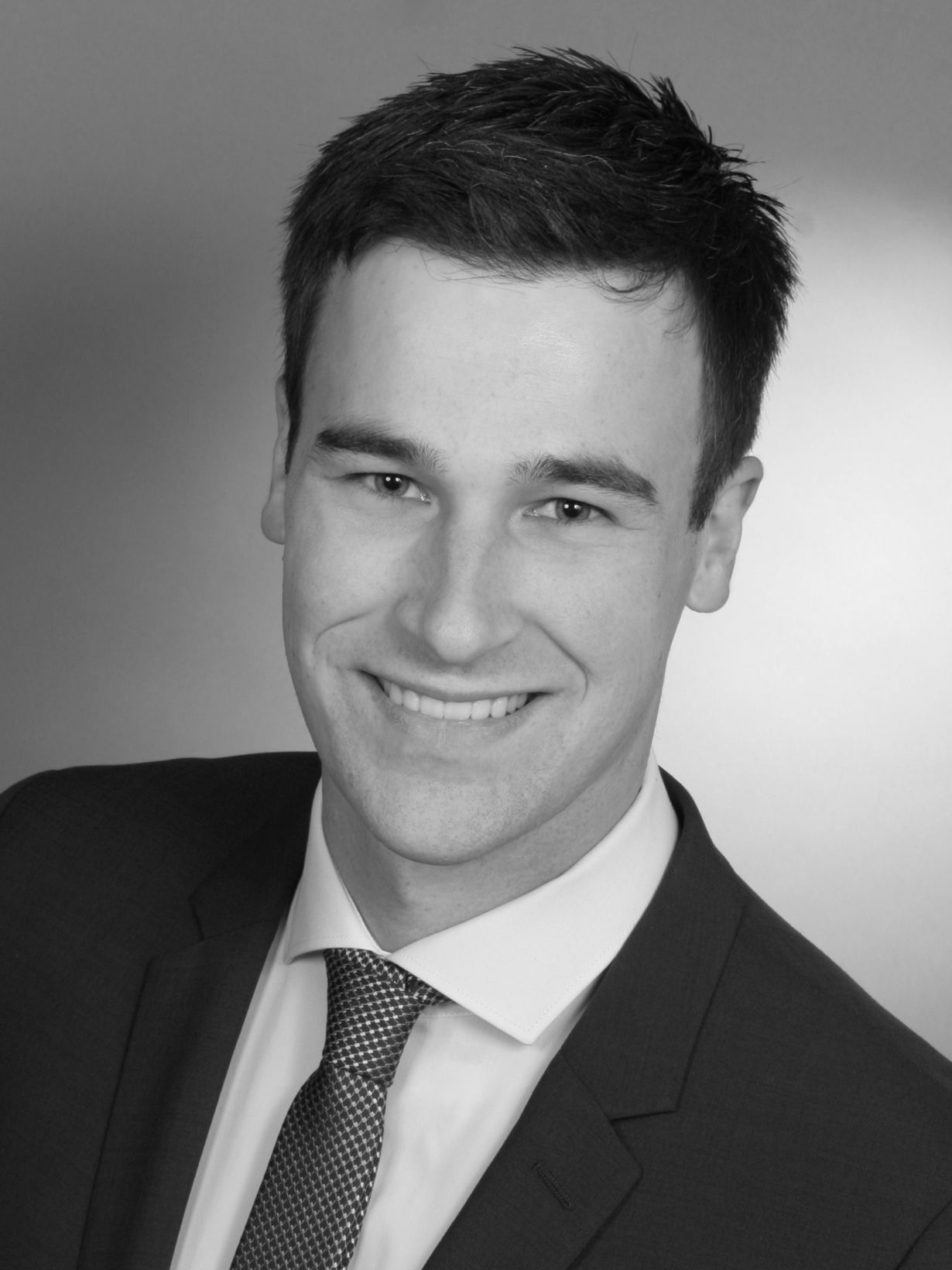}}]{Benedikt Kalde}
received the B.Sc. degree in Computer Science from Paderborn University and the M.Sc. degree in Computer Science from RWTH Aachen University.
In his master thesis with the Chair of Communication and Distributed Systems (COMSYS) he helped conceptualize and analyze \name's initial design and implemented its first prototype.
Today, he evaluates use cases of emerging technology for a consultancy in the aviation industry.
\end{IEEEbiography}

\vspace{-1.5em}
\begin{IEEEbiography}[{\includegraphics[width=1in,height=1.25in,clip,keepaspectratio]{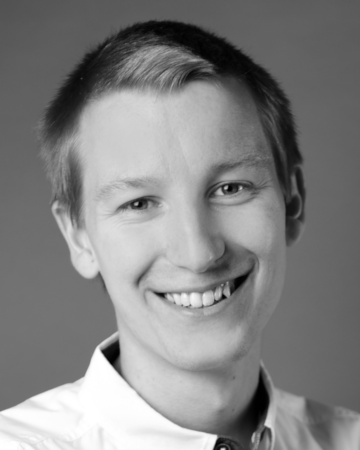}}]{Jan Pennekamp}
received the B.Sc.\ and M.Sc.\ degrees in Computer Science from RWTH Aachen University with honors.
He is a researcher at the Chair of Communication and Distributed Systems (COMSYS) at RWTH Aachen University.
His research focuses on security \& privacy aspects in the Industrial Internet of Things (IIoT).
In particular, his special interests include privacy-enhancing technologies, the design of privacy-preserving protocols, and secure computations as well as their application.
\end{IEEEbiography}

\vspace{-1.5em}
\begin{IEEEbiography}[{\includegraphics[width=1in,height=1.25in,clip,keepaspectratio]{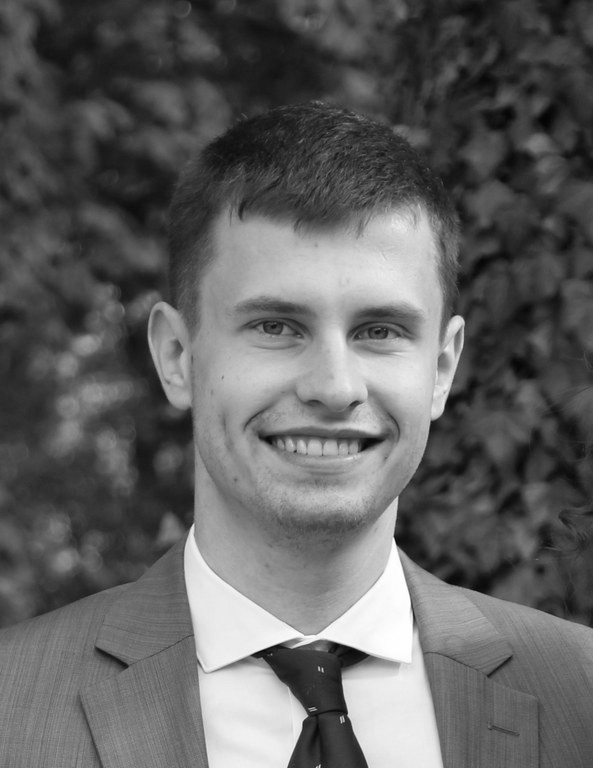}}]{Arthur Drichel}
received the B.Sc.\ and M.Sc.\ degrees in Computer Science from RWTH Aachen University.
He is a researcher at the Research Group IT-Security at RWTH Aachen University.
His research interests include intrusion detection systems, privacy enhancing technologies, and machine learning.
\end{IEEEbiography}

\vspace{-1.5em}
\begin{IEEEbiography}[{\includegraphics[width=1in,height=1.25in,clip,keepaspectratio]{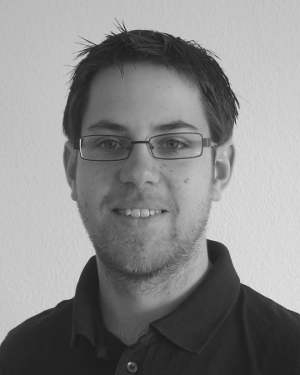}}]{Martin Henze}
received the Diploma (equiv. M.Sc.) and PhD degrees in Computer Science from RWTH Aachen University.
He is a postdoctoral researcher at the Fraunhofer Institute for Communication, Information Processing and Ergonomics FKIE, Germany.
His research interests lie primarily in the area of security and privacy in large-scale communication systems, recently especially focusing on security challenges in the industrial and energy sectors.
\end{IEEEbiography}

\vspace{-1.5em}
\begin{IEEEbiography}[{\includegraphics[width=1in,height=1.25in,clip,keepaspectratio]{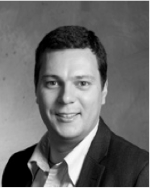}}]{Klaus Wehrle}
received the Diploma (equiv. M.Sc.) and PhD degrees from University of Karlsruhe (now KIT), both with honors.
He is full professor of Computer Science and Head of the Chair of Communication and Distributed Systems (COMSYS) at RWTH Aachen University.
His research interests include (but are not limited to) engineering of networking protocols, (formal) methods for protocol engineering and network analysis, reliable communication software, and all operating system issues of networking.
\end{IEEEbiography}

\end{document}